# Modularity, Architectural Innovation, and New Venture Success[1]

Likun Cao[*], Ziwen Chen[†], James Evans[*‡2]

[*]University of Chicago, [†]Stanford University, [‡]Santa Fe Institute

Startups face a classic dilemma in innovation strategy: should they pursue cumulative, low-risk improvements or disruptive, high-risk breakthroughs? The Henderson and Clark framework suggests that architectural innovation, which reconfigures existing economic modules in novel ways, tends to be disruptive and risky for established organizations, but the success of this strategy for entrepreneurs remains less well understood, largely based on methodological constraints. Building on a complex-economics perspective and advanced computational models, we distinguish architectural innovation from modular innovation, which incrementally updates economic modules, and modular invention, which forges new ones, within the entrepreneurship context. Then we examine how each strategy influences startup performance. We analyze 298,915 U.S. venture-funded start-ups from 1976-2020, embedding company descriptions within a dynamic semantic space constructed from business and patent discourse to measure innovation structure across the entire economy. Event history models reveal that architectural innovation leads to successful IPOs and high-value acquisitions, while both modular innovation and invention increase the risk of failure. By comparing the outcomes of architectural and modular innovation and invention, this paper reveals that what is typically seen as the riskiest form of innovation can, for startups, be the safest route to success. This reconceptualization inverts the trade-off between exploration-exploitation typically assumed in organizational learning with critical implications for entrepreneurial strategy and innovation policy.

**Highlights:**

- Startups face an inverted innovation trade-off compared with established firms
- Architectural innovation is safest for startups while riskiest for incumbents
- Architectural innovation enables rapid experimentation and for new venture success
- Modular innovation and invention bring higher risk of failure for new ventures
- Semantic embedding methods enable economy-wide analysis of innovation strategies

---

[1] We are grateful for support from AFOSR #FA9550-19-1-0354 (to JE), NSF #1829366, and the Institute for Humane Studies Grant IHS017725 (to LC).

[2] Correspondence to James Evans, Knowledge Lab and Sociology, University of Chicago at jevans@uchicago.edu.

# Modularity, Architectural Innovation, and New Venture Success

Startups face a classic dilemma in innovation strategy: should they pursue cumulative, low-risk improvements or disruptive, high-risk breakthroughs? The Henderson and Clark framework suggests that architectural innovation, which reconfigures existing economic modules in novel ways, tends to be disruptive and risky for established organizations, but the success of this strategy for entrepreneurs remains less well understood, largely based on methodological constraints. Building on a complex-economics perspective and advanced computational models, we distinguish architectural innovation from modular innovation, which incrementally updates economic modules, and modular invention, which forges new ones, within the entrepreneurship context. Then we examine how each strategy influences startup performance. We analyze 298,915 U.S. venture-funded start-ups from 1976-2020, embedding company descriptions within a dynamic semantic space constructed from business and patent discourse to measure innovation structure across the entire economy. Event history models reveal that architectural innovation leads to successful IPOs and high-value acquisitions, while both modular innovation and invention increase the risk of failure. By comparing the outcomes of architectural and modular innovation and invention, this paper reveals that what is typically seen as the riskiest form of innovation can, for startups, be the safest route to success. This reconceptualization inverts the trade-off between exploration-exploitation typically assumed in organizational learning with critical implications for entrepreneurial strategy and innovation policy.

**Highlights:**

- Startups face an inverted innovation trade-off compared with established firms
- Architectural innovation is safest for startups while riskiest for incumbents
- Architectural innovation enables rapid experimentation and for new venture success
- Modular innovation and invention bring higher risk of failure for new ventures
- Semantic embedding methods enable economy-wide analysis of innovation strategies

# Introduction

How do organizations emerge to identify, capture, and serve profitable new niches within the economy? This is the central, high-stakes question underlying organizational design and entrepreneurial strategy. A systems perspective suggests that companies and the products and services they produce are constructed from combinations of components interconnected into modules. From this view, innovation can take one of three forms: architectural innovation that reconfigures relationships between existing modules, modular innovation that introduces disconnected but existing modules into an established system (Henderson and Clark, 1990), or something we call modular invention that creates new modules from disparate components which can be subsequently used across the economy.

Henderson and Clark (Albert and Siggelkow, 2024; 1990) developed foundational theory for the strategic trade-off between architectural and modular innovation within established firms. For such firms, modular innovation reflects a safer choice by building incrementally on an existing production function without disrupting its overarching design, whereas architectural innovation poses greater risks by undermining established routines and organizational competencies (Leonard-Barton, 1993; Tushman and Anderson, 1986). As such, Henderson and Clark posit that architectural innovation will typically be left for new entrants to the market.

Yet empirical evidence from this literature has focused almost exclusively on incumbents in manufacturing—automobiles (Abernathy and Clark, 1985; Christensen, 1997), semiconductors (Burgelman, 1994; Henderson and Clark, 1990), air conditioning (Henderson and Cockburn, 1994; Rosenbloom and Christensen, 1994), and similar mature sectors—leaving the optimal strategy for new entrants unresolved. Here we ask: if most successes in architectural innovation come from entrants, do most entrant successes result from architectural innovation? Based on recombinant innovation theory, we theorize that the innovation choices faced by entrepreneurial firms invert those of established firms. While architectural innovation is risky for incumbents, it may represent the safest path for entrepreneurs, enabling rapid experimentation through reconfiguring existing modules within the economy. Conversely, modular innovation, which is safe for established firms, offers no such advantage to start-ups lacking experience or an established production system. We articulate a third natural option common among start-ups: modular invention, where firms develop new technologies designed for use by themselves or others (Cohen et al., 2000). Despite being widely pursued and appearing focused and safe, it involves long-cycle experimentation and often short-term failure.

Testing these hypotheses about entrepreneurial search at scale would be difficult with traditional econometric methods. The nature of entrepreneurial ventures engaged in creative destruction means that their production functions often seek to disrupt existing frameworks. Therefore, entrepreneurial innovations cannot be easily studied using structural approaches like design structure matrices (DSM) and NK simulations, which rely on strong structural assumptions about fixed numbers of combinable components and well-defined system boundaries (Fleming, 2001; Taylor and Greve, 2006).

To date, entrepreneurship scholars have tended to examine contexts of innovation from *whom* ventures obtain knowledge (McDonald and Eisenhardt, 2020; Ter Wal et al., 2016), *how* they attain information (Baker and Nelson, 2005; Gans et al., 2019; Sarasvathy, 2001), and *where* they position themselves (Hannan and Freeman, 1984, 1977; McKnight and Zietsma, 2018), but they have rarely investigated optimal innovation strategies at scale, as modeling such strategies across entrepreneurial ventures cannot be accomplished within existing frameworks.

Given the inherently disruptive nature of entrepreneurship, achieving this goal first requires a framework capable of representing the economy as a whole.

To empirically explore our question, we turn to advanced computational methods. We analyze 298,915 U.S. venture-funded start-ups from 1976-2020, embedding the young company descriptions within a dynamic semantic space constructed from business and patent discourse to measure innovation structure across the entire economy (Aceves and Evans, 2024). This framework advances existing research by going beyond traditional industries and assessing the impact of architectural and modular innovation and invention strategies on entrepreneurial success in both established and nascent markets.

Event history models reveal that architectural innovation leads to successful IPOs and high-value acquisitions, while both modular innovation and invention increase the risk of low-priced acquisitions or closure. To understand the mechanisms underlying these patterns, we conduct mediation analysis examining how innovation strategies affect commercialization speed and investor diversity. We find that architectural innovation accelerates the path from seed to early-round funding by nearly three years and attracts investors with more diverse expertise, while modular invention slows development and concentrates investor backgrounds.

Our findings show how entrepreneurial firms exploit existing economic and technological modules to explore profitable reconfigurations, inverting the exploration-exploitation trade-off typically assumed in organizational learning. This reconceptualization has critical implications for entrepreneurial strategy and innovation policy in both established and emerging sectors, particularly for 21st-century challenges like developing a circular economy and decarbonizing transportation systems (Henderson, 2021).

# Theory and Hypotheses

## The Henderson-Clark Model and Organizational Success

Organizations can be viewed as complex systems that exhibit hierarchical modularity, where subsystems or modules interface to produce emergent capabilities beyond the sum of their parts (Anderson, 1999; Levinthal, 1997; Simon, 1991 [1962]). While this framework has been most classically applied to organizational design, where departments are seen as modules and the organization as the system that couples them (Galbraith, 1973; Sanchez and Mahoney, 1996; Thompson, 1967), it has also been extended to organizational strategies, where organizations are regarded as as systems in which technologies, routines, personnel, and resources are brought together to achieve competitive and profitable outcomes (Ethiraj and Levinthal, 2004; Fleming and Sorenson, 2001; Rivkin and Siggelkow, 2003). Based on this systems perspective, innovation can occur in three ways: (1) by creating new modules, (2) updating systems through incremental module replacement, or (3) modifying the system architecture of connection between modules.

Henderson and Clark conceptualize the latter two strategies as modular and architectural innovation (Henderson and Clark, 1990). Modular innovation occurs when core design concepts and production functions remain stable but one or a few modules are altered without changing the overall system structure. Consider upgrading a car engine while retaining the same drivetrain, frame, suspension, brakes, electrical and fuel systems. Here, one module—e.g., the motor—improves, but its interfaces and relationships with other modules remain unchanged. Architectural innovation, by contrast, involves changes in how modules connect and integrate, while the core design of individual components remains relatively stable. Henderson and Clark illustrate this with the shift from mainframe to personal computers, where most modules, including the processing unit, memory hierarchy, and I/O system remained the same but became reorganized in fundamentally new ways. Beyond these two strategies, a third core strategy involves modular invention that creates new modules (e.g., the invention of lidar or laser-based radar), which can be used by other organizations for their productive systems as well (e.g., lidar for cars with self-driving capabilities). We illustrate these three strategies in Figure 1.

In the top panel of Figure 1, we contrast modular and architectural innovation within the production system of established firms. In the bottom panel, we visualize these strategies within the broader economic context from which new ventures emerge. The *Established Company* in this example uses three modules of three components each, all of which already exist in their mature forms within the economic space. *New Venture A* is focused on modular invention, bringing three economic and technical components into a proposed new module (e.g., lasers, radar, and car navigation). *New Venture B*, instead, engages in modular innovation by using two of the same three modules as the *Established Company*, but replacing a single one (e.g., camera-based navigation with lidar). *New Venture C* undertakes architectural innovation by pulling together preexisting, established modules and combining them in a new way across the economy (e.g., existing content libraries, internet streaming, and recommendation algorithms by digital streaming platforms like Netflix and Spotify).

---

Figure 1 about here

---

Organizations may pursue all three innovation strategies to gain competitive advantage, but their knowledge base and resources limit how effectively they can apply and benefit from each. Studies have identified the boundary between incumbents and new entrants as the critical line shaping the relative returns from architectural and modular innovations. Industrial incumbents, often equipped with deep expertise and refined operational routines, tend to benefit most from incremental, modular innovation. Through improving existing technologies while minimizing disruption, such firms can often outcompete other firms within their established markets (Tushman & Anderson, 1986). Given their reliance on accumulated knowledge and optimized processes, established firms often find architectural innovation—which requires a significant restructuring of internal knowledge frameworks and coordination mechanisms—too costly and disruptive to the existing order (Leonard-Barton, 1992). In contrast, this work characterizes new entrants as unconstrained by legacy architectures and greater flexibility to explore radical and architectural innovation. Without entrenched commitments to existing systems, they can explore novel configurations, challenge industry conventions, and search for new design possibilities with little cost. This adaptability allows them to drive or respond dynamically to disruptive technological paradigms (Christensen, 1997).

## Innovation Strategies of Entrants and Entrepreneurs

Both the Henderson-Clark (1990) and disruptive innovation models (Christensen, 1997) focus primarily on incumbents, framing entrepreneurs' strategies as derivative of those pursued by incumbents. These theories, however, provide little practical guidance for entrepreneurs or theoretical guidance for entrepreneurship scholars, as the strategies least effective for incumbents are not necessarily most effective for other market players. Put differently, even if most successful cases of architectural innovation are associated with new entrants, this does not mean that architectural innovation is a sufficient condition for startup success, nor that it is less risky than alternative innovation paths. The observed association may reflect survivor bias: many startups attempting architectural innovation may have failed and remain unobserved. Guiding principles of organizational innovation for new ventures therefore have yet to be identified.

Entrepreneurs take one of two roles: either entering an established market to challenge incumbents (Christensen, 1997; Henderson and Clark, 1990), or actively defining and shaping a nascent market (McDonald and Eisenhardt, 2020; Sarasvathy, 2001). In the first role, both the Henderson-Clark model and disruptive innovation theory view modular innovation as a problematic strategy for entrants in established markets. Entrants lack the knowledge base and organizational capabilities required to displace incumbents through incremental, modular innovation. Under such circumstances, architectural innovation might be a more feasible path. Without the constraints of legacy designs, entrenched routines, or established supplier relationships, entrants may enjoy greater strategic flexibility to pursue radical experiments and open new market niches (Christensen, 1997). Nevertheless, architectural innovation may also expose entrepreneurs to other challenges, such as skepticism and negative evaluation from critical audiences, including investors and customers, who struggle with how to categorize their novel combinations (Hsu and Hannan, 2005; Navis and Glynn, 2010; Zuckerman, 1999). The relative risks associated with modular and architectural innovation for new entrants have therefore not been systematically assessed. In other words, while modular innovation has been identified as a potential dead end, there is no current evidence to suggest that architectural innovation could not lead to new risks and an equally, or even more, fraught future.

Unlike established markets, nascent markets are viewed as highly uncertain. With limited agreement on the best strategies to pursue there, many studies offer decision-making frameworks rather than specific decision guidance. For instance, Sarasvathy (2001) emphasizes the importance of continuous experimentation and iterative learning to refine an organization’s niche and offerings. Follow-up work provides similar suggestions (Gans et al., 2022; Ries, 2011). These frameworks recognize that nascent markets are dynamic and hard to predict, requiring entrepreneurs to navigate trial and error, rapidly responding to feedback. Discussion of modular and architectural innovation has been largely absent from entrepreneurship discourse in emerging markets, likely because the concepts are typically understood with respect to existing operational systems, and perceived as hard to measure when such stable systems are absent.

Therefore, there is a paucity of direct evidence about entrepreneurial innovation strategies in both established and nascent markets. While this gap can partly be attributed to the lack of a theoretical framework for identifying architectural innovation and modular innovation, especially in the absence of established firms, it also stems from methodological constraints that limit our ability to characterize invention strategies of entrepreneurial firms at scale. For example, modeling entrepreneurial innovation strategies in both established and nascent markets requires a model of the entire economy.

To address this gap, we theorize the role of module-making, replacing, and rewiring for entrepreneurship in the context of a macro economic system. We employ recent AI methods to construct a dynamic map of the economic and technical landscape, from which startups emerge, develop and die, enabling an empirical investigation of the returns to the three innovation strategies in a robust and replicable manner.

## Revisiting Entrepreneurial Innovation with Complex Economic Systems

Although existing theories have explored various entrepreneurial innovation strategies, a comprehensive framework that systematically compares their relative effectiveness across dynamic market contexts remains underdeveloped. Research on complex economic systems offers a useful alternative perspective for theorizing pathways to successful entrepreneurial innovation (Hidalgo et al., 2007; Mealy et al., 2019). Complexity economics views the economy as a complex system in which economic sectors—and their underlying productive functions—serve as interconnected components. The economy as a whole operates through the intricate interactions and combinations of these modular elements (Anderson et al., 1988; Arthur, 1995; Arthur et al., 1997). Although these studies operate at a different analytical level than engineering approaches to complex systems and are therefore less commonly applied by organizational scholars, they nevertheless offer a valuable theoretical framework for entrepreneurship research. As this theoretical framework does not depend on an existing operational system as a reference, it supports the mathematical modeling of innovation strategies within the broader economic system, and helps explain how different strategies alter the distribution of risks and rewards for startup founders. Through this framework, the three distinct forms of entrepreneurial activity can be empirically modeled.

The first and most intuitive strategy involves creating new basic components for the economic system, i.e., inventing new functioning technological modules on which organizations depend. Such inventions address customer needs and expand human possibilities. Technological advances at this foundational level are often achieved by directly controlling, enhancing, and recombining natural capacities, with transformations occurring primarily at the early stage of socio-technical evolution. For instance, Louis Pasteur first controlled the anthrax bacillus in his laboratory before testing the potential for vaccination and ultimately using it to protect French livestock through a large-scale statistical demonstration (Latour, 1993). These inventions laid the groundwork for modern biological and medical technologies, such as advanced vaccines, therapies, and public health protocols, that now underpin the pharmaceutical industry. We refer to these as *modular inventions.* Although inventing a new module is, by definition, a localized activity within a broader operational system or product (Kauffman et al., 2000; Youn et al., 2015), this strategy can nevertheless constitute the core, and sometimes the sole, focus of a startup—particularly among science-based ventures that emerge from university research. Here, the framework of a complex economic system demonstrates its strength, as it enables us to capture the essence of innovation even in the absence of an existing product as a reference.

Recent studies have highlighted the disadvantages faced by young firms engaged in technological invention (Roche, Conti, and Rothaermel 2020). Ventures that pursue excessive technological differentiation may experience limited growth (Guzman and Li, n.d.). Technological inventions introduce risks at multiple levels. These include the possibility of failing to solve the intended problem, failing to generalize a solution beyond its initial context of development and testing, and failing to capture the value introduced by their technology if appropriated or side-stepped by others. These risks arise not only from competitive pressure exerted by incumbents, but also from the inherent instability of the modular invention strategy. For startups lacking a long runway for development and a competitive moat if successful, any of these risks can prove fatal.

Based on that, we propose Hypothesis 1:

*Hypothesis 1: Relative to entrepreneurial firms that engage in architectural and modular innovation, those that engage in modular invention by inventing new technological components, will be less likely to (a) attract new venture capital, and (b) obtain successful IPOs or high-priced acquisitions, but more likely to (c) fire-sale and (4) close.*

Many entrepreneurs pursue a second strategy, *modular innovation*, which does not involve creating new technologies but instead incrementally alters an established system of production by replacing one or a few modules. This might involve, for example, replacing a technology within an existing business model, which aligns with the notion of modular innovation outlined in the Henderson and Clark framework. Entrepreneurs may not only replace a physical technology in the production process, however, but alternatively introduce an established technology into a new market or application context. This likely faces similar challenges as those proposed by the Henderson-Clark Model (1990). When incumbents already dominate the market, their deep knowledge base equips them with stronger absorptive capacities, enabling them to more readily recognize opportunities for technological upgrades within specific business operations or extensions to new market niches.

Finally, within the economic system, entrepreneurs may also pursue *architectural innovation* at a more macro socio-technical level. This involves novel assembly of mature business modules, i.e., technologies and their established applications, across the existing economic landscape. The strategy, often highlighted in journalistic accounts of well-known startups, is difficult to capture with traditional theories or models. Before such ventures emerge, their business models, economic functions, and product forms are often unimaginable. Through this form of innovation, entrepreneurs combine distant economic elements into complex systems that create emergent value—outcomes impossible from any single module alone.

One example of such an interconnected technological system is Uber, which combined (1) smartphone GPS and mobile computing enabling location awareness technology that already existed in phones, mapping apps, and navigation systems; (2) digital payment infrastructure that had been successfully operating for years; (3) telecommunications networks facilitating real-time data exchange; (4) mapping and route optimization software; (5) two-sided marketplace platforms using the matching mechanism pioneered by e-retailers like eBay; (6) reputation and rating systems; and (7) the independent contractor labor model; and (8) the transportation service itself. What made this architectural innovation is that none of these modules were invented by Uber. They all functioned independently in other contexts. The innovation was in the novel linkage architecture. From there, thousands of startups have followed suit with modular innovations. Consider the phrase “this company is the ‘Uber’ of X”, repeated on nearly 200,000 websites and across thousands of start-up descriptions and commentaries, where X represents everything from private jets and houses to mobile gyms, gasoline and garbage (Barns, 2020; Cornelissen and Cholakova, 2021). Originally assembling these modular subsystems in novel ways unleashes new opportunities for growth. Startups engaging architectural innovation generate value that emerges from the complex interactions among existing economic modules.

Incorporating distant economic modules reduces interference from incumbents, who typically remain anchored in their own domains and are reluctant to expand into others. From the perspective of a complex economic system, this strategy also represents a stable pathway for systemic evolution, as it introduces modularity into the system’s structure. In Herbert Simon’s *Architecture of Complexity* (1991 [1962]), he defined modularity as a distinctive structural property of complex systems, in which each module consists of smaller, self-contained units that can themselves be further divided. This structure enables the system to function as a multi-layered, hierarchically organized assembly. He demonstrated that modularity reduces the risk of catastrophic failure: if an evolving architecture collapsed every time a new combination failed, the emergence of a successful, complex object would be improbable. Simon illustrated the underlying logic with watchmaking. Compared to a hypothetical watchmaker who always begins a new watch with its most elemental components, the watchmaker who starts from modular subcomponents does not need to assess the compatibility of lower-level parts and will not lose progress if the assembly fails, assembling only the modules again.

Since Simon, the principle of modularity has also been examined and found present across a wide range of contexts (Baldwin and Clark, 2006; Tseng et al., 2019; Wagner et al., 2016). We argue that for entrepreneurs, who do not assemble a watch from components but instead create new profit cycles from existing technical and economic elements, this logic similarly applies. In the absence of mature precedents, developing new technologies or applications requires intensive exploration, which can slow a venture's ability to capture market share and rapidly exhaust its limited resources. In contrast, startups that build upon existing modules may be better positioned to rapidly generate emergent value. By leveraging mature segments of the economy, these ventures benefit from established infrastructures, accumulated knowledge, and proven operational routines—resources that not only reduce the risk of fundamental failure but also enable more adaptive experimentation and faster scaling. Faster development speed then enhances both the survival and scalability of startups. It alleviates financial pressure and reduces burn rate (Davila et al. 2003), optimizes resource use efficiency (Dencker et al. 2009), and improves adaptability in dynamic markets (Brown and Eisenhardt 1997).

In this sense, modular recombination offers a pragmatic pathway for entrepreneurial innovation, allowing founders to create novelty through new configurations of what already works, rather than through the uncertain pursuit of entirely new technological foundations. Following this logic, we propose Hypothesis 2:

*Hypothesis 2: Relative to entrepreneurial firms that engage in modular innovation and invention, those that engage in architectural innovation by wiring distant economic sectors, will*

*be more likely to (a) attract new venture capital, and (b) obtain successful IPOs or high-priced acquisitions, and less likely to (c) fire-sale and (4) close.*

Therefore, we argue that the optimal innovation strategies of entrepreneurs could largely reverse those of incumbents, guided by a logic that extends beyond market competition. We structure our hypotheses in Figure 2.

Figure 2 about here

## Quantifying Entrepreneurial Innovations with Computational Algorithms

To capture the changing distance between technological components, and construct metrics of three innovation strategies on the system level, we turn to dynamic embedding models. This approach serves as a foundation for measuring innovation strategies in new ventures, as they can be evaluated through a comparison between novel organizational strategies and the contemporaneous socio-economic status quo.

Recent research in finance has demonstrated that word embeddings based on newspapers have accurately captured economic features in the macro environment, and can be used to predict economic fundamentals such as GDP, consumption and employment growth (van Binsbergen et al., n.d.). Word embedding-based measurements have also been widely applied in studies about business niche, market differentiation (Carlson, 2023), and organizational studies more broadly (Aceves and Evans, 2024).

The concept underlying word embedding is straightforward. Through training a machine learning model on text from a given period, its represent words as high-dimensional vectors to enable the measurement of precise semantic distances between them appropriate to that period (Joulin et al., 2016; Mikolov et al., 2013a; Pennington et al., 2014).[3] With sufficient, relevant text, distance between "word vectors" in embedding space serves as a powerful proxy for semantic relationships underlying the referenced concepts. Substantial research has demonstrated the ability of embedding models to replicate even subtle semantic relationships, including analogies and metaphor (Mikolov et al., 2013b), which correspond to subjective understandings held by human agents (Caliskan et al., 2017; Garg et al., 2017; Kozlowski et al., 2019), but also veridical shifts in society, economics, and culture (Garg et al., 2017; Kozlowski et al., 2019). As such, embeddings represent a powerful instrument from which to measure changing semantic associations within an environment.

When applied to a socio-technical corpus comprising both the business press and contemporary technology, word embeddings capture contextual understandings of the economy. This approach allows us to model the dynamic economic environment by training one embedding per period, and tracking changes in distances between market components in the minds of business audiences over time. Building on this, we directly measure each startup's level of architectural and modular innovation or invention by precisely contextualizing their company descriptions within these embeddings, and we estimate their effects on entrepreneurial success. We further elaborate these technical details in the Method section.

## Methods

[3] Word embedding models are sometimes referred to as "low dimensional" relative to the number of words used in text (e.g., 50-100K) because they reduce this *very* high dimensional space of words to a lower dimension. Nevertheless, considered from the perspective of one, two or three dimensional models common in the analysis of innovation and culture, these spaces are *much* more complex, and reproduce much more accurate innovation associations, as demonstrated below.

## Dynamic Word Embeddings Estimate the Business Landscape

Many innovation studies base their measurements of novel emergence on new combinations from existing classification systems, such as patent sub-classes or paper keywords (Fleming, 2001; Taylor and Greve, 2006). These methods rely on the assumption that classification and citation systems are accurate and stable across domains and time. This assumption naturally breaks down in the dynamic market scenario, where new elements emerge every year and industrial patterns can radically transform across a single decade. To account for this dynamism, recent studies take advantage of contemporary natural language processing (NLP) techniques to extract information from technical and business corpora and generate more refined, dynamic measurements (Kaplan and Vakili, 2015; Shiller and Brown, 2019). As the economic and technological environment plays a fundamental role in start-up strategy decision making (Marx et al., 2014), here we first construct a landscape of business and technical concepts through word embedding (with the basic idea illustrated in Figure 3), which provide the context for entrepreneurial innovations and inventions.

Figure 3 about here

Word embeddings produce dense, continuous models that represent words as high-dimensional vectors to enable the measurement of precise semantic distances between them (Joulin et al., 2016; Mikolov et al., 2013a; Pennington et al., 2014).[4] Distance between word vectors in embedding space serves as a proxy for semantic relationships underlying the referenced concepts. Substantial research has demonstrated the ability of embedding models to replicate even subtle semantic relationships, including analogies and metaphor (Mikolov et al., 2013b), which correspond to subjective understandings held by human agents (Caliskan et al., 2017; Garg et al., 2017; Kozlowski et al., 2019), but also vertical shifts in society and culture (Garg et al., 2017; Kozlowski et al., 2019).

[4] Word embedding models are sometimes considered and referred to as "low dimensional" relative to the number of words used in text (e.g., 50,000) because they reduce this *very* high dimensional word space. Nevertheless, considered from the perspective of one, two or three dimensional models common in the analysis of innovation, these spaces are much more complex, and reproduce much more accurate innovation associations, as shown below.

Word embedding algorithms can also be combined with time stamps to model dynamic socio-cultural context. When multi-slice embedding spaces are derived from a temporal corpus, they enable analysis of evolving meanings (Hamilton et al., 2016; Liu et al., 2020; Yao et al., 2018; Zhang et al., 2016). Outputs from these algorithms are time-stamped word vectors that contain the semantic information for that specific period and are comparable across history.

In this project, we adopt a dynamic word embedding method based on joint training of multiple time-sliced embedding space, which allows us to trace the changes within the dynamic economic environment (See Appendix A.2; and Yao et al., 2018). For developing theoretically informed measures, this method even outcompete the recent large language model representations, which rely on imperfectly dated web-scale data and cannot yet be estimated across dynamic periods in the past uncontaminated by shadows of the future (Kozlowski and Evans, 2024; Ludwig et al., 2024).

Our dynamic embedding approach assumes and enforces a smooth transition between word vectors and their associated meanings in adjacent time periods and trains all periods jointly. Specifically, it alters the objective for the joint word embedding to the following:

$$\min_{U(1),\ldots,U(T)} \frac{1}{2}\sum_{t=1}^{T} \|Y(t) = U(t)U(t)^T\|_F^2 + \frac{\lambda}{2}\sum_{t=1}^{T} \|U(t)\|_F^2 + \frac{\tau}{2}\sum_{t=2}^{T} \|U(t-1) = U(t)\|_F^2 \quad (1)$$

The first term in this objective, the $L_2$ norm of the difference between the empirical word proximities $Y(t)$ and the modeled proximities $U(t)$, represents the standard word embedding objective. This objective guides optimization of the algorithm to accurately model word distances within time periods, and the $L_2$ norm regularizer is used, as in ridge regression, to improve out-of-sample predictive performance of the model. Dynamic word embeddings add two additional terms that: 1) allow the algorithm to minimize unnecessary word vector loadings (the $L_2$ norm of $U(t)$), and 2) shrink the distance between time-adjacent embeddings (the $L_2$ norm of the difference between $U(t-1)$ and $U(t)$).[5] In minimizing this cost function, adjacent word embedding spaces are grown in alignment, with a $t \times n \times k$ matrix, where *t* is the number of time slices, *n* is the number of words, and *k* is the dimensions of word vectors. The business landscape of year *t* is stored in one $n \times k$ slice, which is optimized to remain similar to the *t*-1 and *t*+1matrices. In this way, the jointly trained algorithm retains words that newly appear in the corpus over time and keeps adjacent semantic spaces close in structure. This allows us to construct comparable and historically accurate representations of the structure of concepts within business discourse for each year over the nearly half century we examine.

We use two datasets as input to the dynamic word embedding model: (1) the ProQuest business news, and (2) the USPTO full text patent corpus, 1976-2020. ProQuest business news includes 108 English newspaper and magazine full-text articles, published in the U.S. and falling into the category of "Business and Economics". This contains major business publications, including the *Wall Street Journal*, *Bloomberg BusinessWeek*, *the Economist*, and *American Banker*, among many others. It provides temporal information about relationships between business concepts, including both technologies and their application areas, based on how reportage discussed them in the context of contemporary and imagined future business. In total, our dataset includes over 6 million business news records (1976-2020; see Appendix A.1 for additional corpus details). We complement this dataset with expert knowledge about the nature of new venture technologies and their applications, incorporating full text descriptions from the USPTO patent dataset (1976-2020). We adjust the weights between news and patents to keep application and technology elements balanced during the training process. This allowed us to represent both business discourse and the technological knowledge basis underlying U.S. venture creation across this period.

[5] This dynamic word embedding approach relies on the formal equivalence between the standard negative sampling approach to word embedding optimization used by `word2vec` and the low-rank factorization of a pointwise mutual information (PMI) matrix of word vectors (Levy and Goldberg, n.d., 2014), and uses the PMI as input to the objective above.

The output of our dynamic word embedding is a 45-slice word embedding space, modeling our joint business-technology corpus from 1976-2020. Each slice is made up of 50-dimensional vectors for 227,714 words, representing the geometric distribution of business concepts (i.e., words) from a given year. It provides an intuitive and accurate model of the changing market, faced by all new ventures.

We demonstrate the strengths of our dynamic embedding models through several intuitive examples. When the business functions of a word change, its vector drifts across the landscape. In Figure 4, we use “Amazon” as an example to show this dynamic. From 1993 to 2017, Amazon transformed, at first dramatically from the Brazilian rainforest surrounded by concepts “tropical”, “forest” and “farmland”, to an internet retailer in 2001 nearest “bookseller”, “etoys” and “priceline”, other online retailers of toys and tickets. Thereafter Amazon continued to evolve. In 2006, the year after its introduction of Prime video and music, it was closer to “youtube”, “video sharing”, and the social media platform “myspace”. With the roll-out of its Fire Phone in 2014, following the success of its Kindle Fire reading and internet devices, Amazon moved closer to “whatsapp”, “smartphone”, “samsung" and “sony”. The emergence of Amazon as the hegemonic retailer it is today, covering all product categories with advertising revenue reaching the billions in 2017, positioned the company near “online”, “advertising”, “ecommerce”, and “facebook”. Other cases of the evolution of company positioning for Apple, Tesla, and Theranos are provided in the Appendix A.2. For example, with the dramatic fall of Theranos in 2016, words surrounding the company before and after dramatically diverged in sentiment. Also see Appendix A.2 for additional validations of our evolving business landscape.

---

Figure 4 about here

---

## Mapping New Venture Descriptions within the Evolving Business Space

The dynamic space provides an opportunity to empirically examine the innovation strategies of new ventures within their socio-historical contexts. In this study, we use two databases to independently assess the position of new venture strategies within the evolving semantic space of business described above. First, we use CrunchBase (CB), a commercial database serving entrepreneurs and investors (N=298,915 by 2020). Second, for independent validation, we use VentureXpert (VX) from Thomson Reuters (N=63,492 by 2018). Both datasets contain information regarding U.S. companies established 1950 to present. Both datasets provide a long description for each company, with detailed information about its major markets, relevant technologies, and strategies. Meanwhile, these two descriptions are developed independently: in Crunchbase, these descriptions are drafted by representatives of the company itself and provided to the database, which has been widely used in strategy-related studies (Chae and Olson, 2021; Savin et al., 2022). In VentureXpert, analysts from Thomson Reuters independently construct descriptions of comparable length and format. We reported model results based on CrunchBase in the main text, and repeated all analyses with VentureXpert as robustness tests, presented in the Appendix D.2.

We project company descriptions, either from Crunchbase or VentureXpert, onto our evolving embedding space.We first identify all vectors related to words in the business descriptions. Then, we construct measures for architectural innovation, modular innovation and modular invention, based on the geometric relationship between these word vectors in the semantic space.

## Constructing Measurements for the Three Innovation Strategies

In this section, we build separate measures for architectural innovation, modular innovation, and modular invention. Given the importance of module boundaries in distinguishing the three strategies, we first identify major economic modules in the market from the structure of the semantic space. Words cluster closely within the embedding space only when they are widely used together in business and technological discourse. In this way, clustering algorithms capture well-established modules in the shared knowledge of the economic system.

We use a high-performance topic identification method, the identification of "discourse atoms" that draws directly upon our word embeddings (Arora et al., n.d.; Arseniev-Koehler et al., 2022), to partition the semantic space into coherent semantic regions. This algorithm uses word embedding vectors as inputs, and partition words into *k* "discourse atoms" that tile the semantic space into coherent regions. We identify 200 atoms for each time slice to balance human interpretation and information loss (Choi et al., 2021; Kaplan and Vakili, 2015), but our findings are insensitive to this number or clustering method (see Appendix C). In table 1, we use 1990 and 2010 data as an example to show the performance of discourse atom clustering.

Table 1 about here

The word embedding space and discourse atoms provide the basis for us to construct three measures that robustly distinguish architectural innovation, modular innovation, and modular invention.

***Modular Invention***: We measure the average distance between company components within the same discourse region to directly measure modular invention. A large distance within the same region indicates that the new venture is attempting to assemble economic and technological elements that have not yet matured into coherent economic modules. A small distance, by contrast, suggests that clusters of their components have been frequently used together in the past, reflecting their recognized modularity within the marketplace. Less modular invention often liberates a new venture to engage in other types of higher-order modular or architectural innovation.

***Architectural Innovation***: Next, we measure the average cosine distance between a company's relevant modules to establish the surprise associated with innovations in combination architecture. High average distance between a company's modules indicates that this architectural connection has not yet been actively discussed, developed or deployed together in industry. Low average distance between modules suggests that a firm has brought together modules frequently considered together.

***Modular Innovation:*** Finally, to capture modular innovation, we measure the standard deviation of cosine distances among a company's relevant modules. A small value indicates that the modules are similarly distant from one another, whereas a large value suggests that some modules are closely related while others are far removed. This pattern signals that only one or a few modules have been replaced by distant ones, thereby incrementally altering common production functions, as articulated by the Henderson-Clark model. We illustrate these three different innovation strategies in Figure 5.

## Variable Construction

***Independent Variables: New Venture Innovation Strategies***

We measure architectural innovation, modular innovation, and modular invention using geometric distances within the semantic space, as described in the previous section.

***Dependent Variable: New Venture Performance***

Following prior research, we used the milestone events in new ventures' growth trajectories as dependent variables. CrunchBase and VentureXpert both record major events in company lifecycles, including founding, obtaining new funds from venture capital firms, acquisition, initial public offering (IPO), and closure. Each of these events can be taken as a milestone that reveals the competitive strength or weakness of an individual company, and the success or failure of its investors (Ljungqvist and Hochberg, 2009).

We treat IPO and high price acquisitions as proxies for successful growth and development (Chang, 2004; Keyhani et al., 2022). Crunchbase contains records for both types of events. We identified M&A transactions with the top 30% price in each industry as high-priced acquisitions, with dollar values deflated by the Consumer Price Index (CPI). New ventures reporting an acquisition, but with missing prices, are assigned to the middle-low priced acquisition classification; a strategy guaranteeing high precision for successful outcomes (i.e., all acquisitions classified as high-priced are verified) at the cost of lower recall. In addition to IPO and M&A, we also take getting new funding from venture capital firms as a positive sign of venture development (Roche et al., 2020) and bankruptcy as a sign of failure.

We treat these four types of events (IPO/high price acquisition, receiving new funding, middle-low price acquisition and closure) as competing outcomes for a venture in any given period. In the Appendix D.4, we also re-estimate our models with two alternative codings (combining all acquisitions into the same category and separating high-price acquisition and IPO into two independent categories) with the results remaining robust. A company can have several observations in the data, reflecting different developmental stages. We take IPO and closing as final exits, however, and all events after final exit (like revival after closing) are ignored.

***Controls***

Control variables include (1) text features associated with the business description, including text length (numerical), the existence of very rare words (dummy), and lack of technical words (dummy); (2) features of the company, including funding history (number of historically accepted funding rounds before the observation period), founders' identity (whether the founder is a woman or minority), and company's media usage (whether the company uses Facebook and/or Twitter platforms); (3) the market environment in the year the venture was founded, measured by the growth rate of companies in the same industry; (4) temporal-spatial features, including year of observation, and geographical region of company headquarters; and (5) perceived familiarity of technology components as a function of frequent, recent usage (Fleming, 2001). We incorporate this last technology familiarity variable to capture the recent frequency of concept usage in the entrepreneurship community alone.

***Robustness Tests on Measurement Construction***

We note that new ventures may change over time. Nevertheless, once companies mature such that they enter Crunchbase or VentureXpert, their descriptions rarely change. In order to assess the robustness of our findings to strategic pivoting among new ventures (Kiss and Barr, 2015), we collected historical records of the CrunchBase dataset from the Wayback Machine, a non-profit internet archive that preserves historical records of persistent websites. This record allows us to trace changes in company descriptions on a monthly basis. We carry out validation analysis based on this longitudinal data. All results reported below remain robust and do not change with time-shifting descriptions (see Appendix D.3).

## Model Specification: Event History Model with Competing Risks

We adopt a multi-outcome survival model to explore how innovation strategies shape new venture futures. As many outcomes are mutually exclusive—a venture cannot go IPO and bankrupt at the same time—we avoid the commonly used Cox model, and use models that deal with competing risks. For this consideration, we apply the cause-specific flexible parametric survival model (FPM) developed by Mozumder and colleagues (Mozumder et al., 2017). This model identifies the factors that lead to a specific outcome, taking all other possible outcomes into consideration. We use the cause-specific Cox model as a robustness check. The results remain robust and do not change our main conclusions (see Appendix D.1).

## Structure Equation Models to Explore Mechanisms

Our final analysis explores mechanisms underlying our findings. For this purpose, we construct a mediation analysis. We take each new venture as a single observation and set final outcomes (IPO and high-priced acquisition, funding, low-priced acquisition, and closure) as dependent variables. The two types of invention strategies serve as major predictors, and we construct two mediators based on theoretical discussions of modularity: time to market, and VC diversity. If the logic of modularity holds—namely, architectural innovation enables firms to leverage existing economic modules, we would expect such firms to exhibit faster growth, as they are buffered against the risks of system collapse. We would also expect them to attract a broader and more supportive investor base, as they are built upon stable and widely recognized modules from different perspectives within the economy. Both patterns follow naturally from the core logic of the modularity framework.

***Time to Market.*** We measure the growth speed of development with the time difference between seed and early round (A or B round) investment. In the seed funding stage, new ventures typically have begun to pursue their ideas, but remain unproven within the market. By contrast, early round funding occurs when product-market fit has been established and the team is preparing to scale. Thus, the time difference between the two rounds proxies for the growth speed of a new venture.

***VC Diversity.*** In order to trace the breadth of expertise and background mentoring a firm, we trace past investment records for each VC firm when a new deal occurs. These include a terminology of Crunchbase industry keywords. The difference between VC backgrounds is calculated as pairwise Jaccard diversity (i.e., 1-Jaccard similarity) for industrial expertise keywords, and the diversity of VC background for one transaction is the average of all pairwise Jaccard diversity for participating VC firms.

As our two mediators are constructed based on information about seed and early rounds (A&B rounds) of funding, in our mediation analysis we only include startups with clear information about these investments. Thus, the sample size for our mediation analysis is markedly smaller than the event history analysis (N=49,304), with an even smaller sample for robustness testing (N=19,979). Results show very similar patterns with our main models and support our hypotheses.

## Overview of the Research Design

Figure 6 reveals our analysis workflow, computational tools and datasets for each step. We perform the dynamic word embedding and discourse atom modeling in Python 3, and complete event history and mediation analysis in Stata 18.

Figure 6 about here

# Results

## Descriptive Statistics

Descriptive statistics of our major variables are reported in table 2. The two databases, Crunchbase and VentureXpert, generate similar distributions for most variables. The only noticeable difference is that VentureXpert, on average, has higher numbers for fund-raising and IPOs. This could be because Crunchbase, which mainly relies on volunteer reporting and automatic information-scraping, has provided more up-to-date self-reporting and broader sample coverage, including information on smaller organizations with limited funding. Statistics in Table 2 guarantee that the two databases are comparable.

Table 2 about here

We also report a correlation table of the variables in Table 3.

Table 3 about here

We initially explore the relationship between IVs and DVs with a simple descriptive analysis. We transform numerical predictors into ten equal-size quantile groups as the *x*-axis, and calculate the average proportion for IPO, M&A and closure (or average number of new funding rounds) for each group as the *y*-axis. Evidence from Figure 7 supports our hypotheses 1 and 2: In general, architectural innovation has a positive correlation with IPO/high price acquisition and new funding attainment, and a negative relationship with other acquisitions and closure. Meanwhile, modular inventions are associated with less fundings and higher probability of low-price sale.

Figure 7 about here

## Event History Model

These preliminary results are confirmed with event history models, reported in Table 4. After controlling for all covariates and accounting for competing risks, the relationships between the focal independent and dependent variables become clearer. When new ventures engage intensively in modular invention, it is associated with a lower probability of IPO/high price acquisition and new funding. Specifically, we observe a coefficient of -0.939 for local distance (modular invention) on the logged hazard rate that a company will achieve IPO or high priced acquisition. This suggests that with a maximum increase in local distance (modular invention), the hazard of going public or being richly bought out decreases by nearly 60.9% ($0.609 \approx 1-e^{-.939}$).[6] Similarly, a maximum increase in local distance leads to a 34.5% ($1-e^{-.424}$) decrease in the hazard of obtaining new funding. When a new venture engages in modular invention, its likelihood of ultimate success drops substantially.

Interestingly, when considering competing risks, modular invention (local distance) also increases the hazard of middle and low priced acquisition by 67.6% ($e^{.517}-1$). This suggests that modular invention, although not the optimal strategy for a founding team, can still make the new

[6] Technically, our model yields the cause-specific sub-distribution hazard rate, which was designed in order to analyze competing risks between different events, as we do here.

venture a desirable target for other enterprises through acquisition.

As predicted by the Henderson-Clark model, modular innovation can be an especially bad strategy for new entrants: a one-unit increase in global standard deviation (modular innovation) is associated with a 61.4% ($1-e^{-.952}$) decrease in the hazard of obtaining new funding, a 37.8% ($1-e^{-.474}$) decrease in the hazard of being acquired even at a low price, and 7.757 ($e^{2.170}-1$) times higher likelihood of closure. Relying solely on modular innovation, a startup cannot create a new market niche and must compete in established markets where incumbents hold overwhelming advantages. Consequently, such ventures are unlikely to be acquired and are most likely to be forced out.

Finally, our models suggest that architectural innovation is the optimal strategy for entrepreneurs. A one-unit increase in global distance (architectural innovation) multiplies the hazard of IPO or high-priced acquisition by 9.56 ($e^{2.258}$), yielding a 856% increase. It also raises the hazard of obtaining new funds by 206.8% ($e^{1.121}-1$). By contrast, the same decreases the hazard of ending in a closure by 94.1% ($1-e^{-2.837}$).

To test the quality and goodness-of-fit of our models, we introduce two measurements from literature: the $\rho_k^2$ measure proposed by O'Quigley and colleagues (O'Quigley et al., 2005), and a more conservative improvement on this measure by Royston (2006). The first is defined as $\rho_k^2 = 1 - exp\left(-\frac{X^2}{e}\right)$, where $X^2 = 2\left(l_{\hat{\beta}} - l_0\right)$ where $e$ equals the number of uncensored observations. The second is defined as $R^2 = \frac{\rho_k^2}{\rho_k^2 + (\pi 2/6) * (1-\rho_k^2)}$, designed to measure the explained variation for event history models. Our models perform well according to these goodness-of-fit measures, with $\rho_k^2$ between 0.39 and 0.40, and $R^2$ between 0.28 and 0.29, across all estimations.

---

Table 4 about here

In Figure7, we estimate the hazard rate for three hypothetical companies on the basis of our model. We assume three "ideal types", with the virtual companies having highest empirically observed scores of architectural innovation (global distance), modular innovation (global std), and modular invention (local distance), respectively.  Insofar as our hypotheses hold, we expect much better outcomes to befall the enterprise exclusively focusing on architectural innovation and worse outcomes for those exclusively investing time and energy in modular invention. Results from Figure 8 meet these expectations. We observe a clear divergence between hazard rates for the three ideal-type ventures. Firms engaged in modular invention are much less likely to receive new funding and achieve an IPO and are much more likely to be sold for a middle to low price. Startups engaged in modular innovation are the most likely to close.

Figure 8 about here

In order to assess the robustness of our event historical findings, we rerun the models using data from VentureXpert. This replicates the model as in Table 4 on an overlapping but largely distinct population of new ventures and events. We further evaluate robustness by accounting for the potential for new ventures to evolve and change with new opportunities in the marketplace. For this, we utilized the Wayback Machine data to account for strategic pivoting in new ventures away from their initial business description and plan. Our results remain robust to these major perturbations and do not change our conclusions (see Appendix D.3).

## Mediation Analysis

Our event history model shows that architectural innovation is associated with new ventures achieving milestone events, while modular invention harms them. But why is it the case? In this section, we conduct a mediation analysis to illuminate mechanisms through which the two approaches influence performance of new ventures through commercialization speed and investment diversity, as suggested by the modularity framework within complex systems.

We first regress the two mediators on our main independent variables (Table 5), and then use the generalized structural equation modeling to estimate and display the effects of two mediators (Table 7) between IVs and DVs. In Table 7, we decompose that effect into direct and indirect effects on startup final outcomes. We illustrate the path for indirect effects in Figure 8.

---

Table 5 about here

---

Table 5 shows the effects of invention strategies on the mediation variables. Results are consistent with our expectations. Holding other variables constant, changing modular invention score (local distance) from the empirical minimum to maximum adds approximately 13.22 months to the period between seed and early round investment (A/B), while changing architectural innovation (global distance) from its empirical minimum to maximum reduces that period by approximately 34.66 months. Higher architectural innovation and lower modular invention is also associated with the likelihood new ventures obtain diverse support from the investor community, as they attract investors with more distinctive professional backgrounds. Increasing architectural innovation (global distance) from its empirical minimum to maximum increases the Jaccard diversity score of VC backgrounds by nearly 0.134, and increasing modular invention (local distance) from its empirical minimum to maximum decreases the Jaccard diversity score of VC backgrounds by 0.104. Given the Jaccard diversity score falls in the range [0,1], a growth of 0.134 represents more than an eighth of the total range, suggesting substantial increase in VC knowledge backgrounds. When we replicate these analyses on all samples with both seed round records and early round records, the pattern remains robust.

After demonstrating the validity of mediators, we add them into generalized structural equation modeling to identify the path of effects. Results are presented in Table 6.

---

Table 6 about here

---

As shown in Table 6, shorter time to receive early-round funding and more diversified venture capital backgrounds both contribute to the higher probability of IPO/acquisition, partly explaining why different innovation strategies lead to fundamentally different outcomes. Overall, the model demonstrates high goodness-of-fit, indicating that it captures a substantial share of the effect pathways[7]. In structural equation modeling, when path A→C co-exists with A→B→C, the total effect from A to C = ac(direct)+ab×bc(indirect). Table 7 shows direct, indirect and total effects between our major variables, and we visualize these effect paths in Figure 9.

---

Figure 9 about here

---

To summarize, our estimated mediation effects confirm our expectations consistently for all three outcome variables. Architectural innovation significantly accelerates new venture development from idea to market. Higher architectural innovation (global distance) and lower modular invention (local distance) also bring in broader expertise and insight from investors, and expand the knowledge base accessible to the founding team. The direct and indirect effects illustrated in Figure 7 echo these patterns, demonstrating that insights from complex economic theory can be effectively applied to the case of entrepreneurship, and that entrepreneurial strategies closely align with the prediction of system evolution.

[7] For goodness-of-fit measures for GSEM, we report AIC and BIC#. To achieve an approximate estimate, we recode the outcome variables as binaries (instead of multi-category outcomes) and re-estimate the equation with an ordinary SEM of the same structure. The output model has an AIC of 784889 and BIC of 786509, compared to 597720 and 599296 for the GSEM. This demonstrates that the GSEM improves in its fit of the current data. The ordinary SEM has a RMSEA of 0.028 (<0.05), CFI of 0.99 (>0.95) and SRMR of 0.001 (<0.08), which means our model has fit the data well and captured variation suggesting that it has captured an additional aspect of the social process. Nevertheless, due to the large sample size, the $\chi 2$ test does not work well with a $p < 0.05$.

Table 7 about here

# Discussion

In this study, we compare entrepreneurship strategies of innovation, and how successful new ventures rely on modularity to hedge against the risks of exploratory failure (Simon, 1991). Based on the economic complexity perspective and its application in organizational innovation with the Henderson Clark model, we hypothesized that when new companies engage in architectural innovation and bridge across remote economic modules, they are more likely to achieve long term success. In contrast, when they engage in modular innovation or invention, crafting new technologies or applying them to new economic problems, they are unlikely to succeed in the marketplace under the time constraints imposed by venture capital and competition.

Drawing upon computational tools, we constructed an evolving map of techno-economic discourse, and used it to anchor our analysis of the modularity and combinatorial novelty of U.S. entrepreneurial ventures (1976-2020). We projected new venture descriptions (298,915 from Crunchbase; 63,492 from VentureXpert) within this map to evaluate the degree to which they engaged in different types of innovation activities.

By testing the relationship between innovation strategies and liquidity events, our empirical evidence demonstrates that the strategic options faced by startups are the inverse of those facing established firms. For the startups, developing or applying technologies in novel ways can involve higher risks and slows new venture development. In other words, modular inventions can be highly risky for new ventures. It decreases the probability that new ventures receive new funding rounds from venture capital firms, achieve an IPO, or sell for a strong price, and increases the probability that they will shutter. In contrast, those engaged in architectural innovation, assembling mature, applied modules into new value-generating systems, were much more likely to obtain rapid new rounds of funding and achieve outsized success through IPO or high-priced acquisition.

Our work directly contributes to the organizational learning and positioning literature in entrepreneurship. Existing research has extensively examined the optimal contexts for organizational learning, including the most effective sources of knowledge (Argote and Miron-Spektor, 2011; Cohen and Levinthal, 1990; Shane, 2000) and the best strategies to learn (Katila and Ahuja, 2002; March, 1991). Research on firm positioning has explored the notion of cultivating an industrial “niche” (Carroll 1985; Hannan and Freeman 1977) and achieving “optimal distinctiveness” (McKnight and Zietsma 2018). Our study contributes to this prior work by offering a new perspective on the optimal innovation strategy for start-ups. Using a novel computational mapping of the economic landscape, we develop measures of architectural innovation, modular innovation, and modular invention, operationalizing concepts from the Henderson–Clark model for the first time in entrepreneurship research. Our empirical analysis identifies architectural innovation as the most effective strategy for entrepreneurial ventures in nascent or established markets, thereby extending and complementing existing debates.

For entrepreneurship scholarship, our approach also helps explain recent empirical evidence about new venture strategy, including the disadvantage of young companies engaged in technological invention (Roche et al., 2020) and the low growth rate of new ventures with too much differentiation (Guzman and Li, n.d.)

On the analytical level of the socio-economic system, our work highlights boundary conditions for the benefits of recombination. Social domains with a strong incentive for innovative speed, such as entrepreneurship, will not typically benefit from modular invention that requires intensive resource inputs and risk. For such missions, innovators can scaffold their search with previously applied technologies. This pattern is evident in the case of the U.S. defense advanced research projects agency (DARPA), which catalyzed the internet and many forms of modern artificial intelligence, including the technology underlying Siri and Alexa. DARPA articulates a funding strategy that coincides closely with architectural innovation. It frequently supports attempts to rapidly assemble diverse, stable technologies into novel systems (e.g., the Internet) designed to surprise rather than be surprised by competitors.[8] Other domains without the same capitalist or military pressure for rapid success, such as science, typically rely less on the modularity of applied technologies, or rely on modularity at different levels of the system. Therefore, while for-profit investors such as venture capital firms may be motivated to support projects centered on architectural innovation, public funding should prioritize more fundamental discoveries and inventions. Although these endeavors carry substantial risk, they generate public goods that enable successful entrepreneurs to engage in architectural innovation (Jones and Summers, 2020).

[8] "The genesis of…DARPA…dates to the launch of Sputnik in 1957, and a commitment by the United States that, from that time forward, it would be the initiator and not the victim of strategic technological surprises" (https://www.darpa.mil/about-us/about-darpa, accessed June 10, 2023).

Finally, our work speaks to theory about the importance of diversity in teams, networks, and organizations for innovation (Page, 2019, 2008). Most previous work on creativity and novelty emergence demonstrates how social environments desirable for innovation catalyze "creative conflicts" (Cao et al., 2022), either by unleashing the free flow of new ideas (Stark, 2011) or by creating a stable environment that supports higher levels of cognitive heterogeneity and resulting debate (Goncalo et al., 2015). Nevertheless, recent work on information benefits of social networks highlight the danger of information overload (Ter Wal et al., 2016). Without a shared cognitive schema, diversity can bring more chaos than creativity. Our work shows that successful recombination mobilizes the modular structure of the knowledge system. Insofar as problems and solutions become stable modules within a complex organization, like technologies and applications in the new ventures examined in our study, the diversity and even incommensurability of those modules may enable that organization to fluidly reorganize in response to new challenges and opportunities.

Despite its merits, our work has multiple limitations. First, we did not have the data to incorporate network effects or geographical considerations into our analysis, despite their proven significance for entrepreneurial success (Sorenson, 2018). Future study could examine spatio-temporal contexts to investigate how such factors moderate the relationship between new venture structure and efficiency. Furthermore, although we have incorporated wayback machine data for robustness checking, it only captures description changes after 2013, and does not allow us to comprehensively examine the full history of new venture pivots. Prior research indicates that when longitudinal data is involved, conclusions drawn from VentureXpert remain robust (Shi et al., 2017). Others have noted that core business ideas typically remain stable as start-ups grow (Kaplan et al., 2009) despite frameworks that encourage them to actively experiment and shift on a tactical level (Gans et al., 2022; Ries, 2011). Our robustness check on the Crunchbase descriptions based on Wayback machine data suggests the same pattern. Therefore, we assume that if we could incorporate complete longitudinal data, our main conclusion would remains robust. More complete data will be required, however, in order to completely rule out a potential interaction between modular company descriptions and company pivots.

Our study highlights that not all inventive combinations are equal. By examining tens of thousands of U.S. venture-funded start-ups over the past half century, we uncover the importance of recombining modular applied technologies to rapidly assemble innovative and persistently value-generating business organizations. Future research should consider how functional modularity scaffolds innovation in fields ranging from science and technology to finance and business to culture and the arts (Wimsatt, 2014). Such work should also look beyond "modularity" (Simon, 1991) to additional principles through which complex social systems hedge against exploratory failure, such as search along "adjacent possible" innovations one-step from the current frontier (Kauffman, 2002). By uncovering the syntax of innovation, such investigations will not only enable deeper understanding of how actors manage risks and opportunities underlying attempted innovations, they will also enable us to anticipate the range of innovations and inventions most likely to emerge.

# References

Abernathy, W.J., Clark, K.B., 1985. Innovation: Mapping the winds of creative destruction. Res. Policy 14, 3–22.

Aceves, P., Evans, J.A., 2024. Mobilizing Conceptual Spaces: How Word Embedding Models Can Inform Measurement and Theory Within Organization Science. Organization Science. https://doi.org/10.1287/orsc.2023.1686

Albert, D., Siggelkow, N., 2024. Architectural Search and Innovation.
Anderson, P., 1999. Perspective: Complexity Theory and Organization Science. Organization Science 10, 216–232.
Anderson, P.W., Arrow, K., Pines, D., 1988. The Economy As An Evolving Complex System: The Proceedings of the Evolutionary Paths of the Global Economy Workshop, Held September, 1987 in Santa Fe, New Mexico, 1st edition. ed. CRC Press, Boca Raton, FL.
Argote, L., Miron-Spektor, E., 2011. Organizational learning: From experience to knowledge. Organ. Sci. 22, 1123–1137.
Arora, Li, Liang, Ma, Risteski, n.d. Linear algebraic structure of word senses, with applications to polysemy. Int. J. Small Craft Technol. https://doi.org/10.1162/tacl_a_00034/43451
Arseniev-Koehler, A., Cochran, S.D., Mays, V.M., Chang, K.-W., Foster, J.G., 2022. Integrating topic modeling and word embedding to characterize violent deaths. Proc. Natl. Acad. Sci. U. S. A. 119, e2108801119.
Arthur, W.B., 1995. Complexity in economic and financial markets. Complexity 1, 20–25.
Arthur, W.B., Durlauf, S.N., Lane, D. (Eds.), 1997. The Economy As An Evolving Complex System II, 1st edition. ed. CRC Press, Boca Raton, FL.
Baker, T., Nelson, R.E., 2005. Creating Something from Nothing: Resource Construction through Entrepreneurial Bricolage. Adm. Sci. Q. 50, 329–366.
Baldwin, C.Y., Clark, K.B., 2006. Modularity in the Design of Complex Engineering Systems, in: Braha, D., Minai, A.A., Bar-Yam, Y. (Eds.), Complex Engineered Systems: Science Meets Technology. Springer, Berlin, Heidelberg, pp. 175–205.
Barns, S., 2020. The Uberisation of Everything, in: Barns, S. (Ed.), Platform Urbanism: Negotiating Platform Ecosystems in Connected Cities. Springer Singapore, Singapore, pp. 79–98.
Burgelman, R.A., 1994. Fading memories: A process theory of strategic business exit in dynamic environments. Adm. Sci. Q. 39, 24.
Caliskan, A., Bryson, J.J., Narayanan, A., 2017. Semantics derived automatically from language corpora contain human-like biases. Science 356, 183–186.
Cao, L., Chen, Z., Evans, J., 2022. Destructive creation, creative destruction, and the paradox of innovation science. Sociol. Compass 16. https://doi.org/10.1111/soc4.13043
Carlson, N.A., 2023. Differentiation in microenterprises. Strategic Manage. J. 44, 1141–1167.
Chae, B., Olson, D.L., 2021. Discovering Latent Topics of Digital Technologies From Venture Activities Using Structural Topic Modeling. IEEE Transactions on Computational Social Systems. https://doi.org/10.1109/tcss.2021.3085715
Chang, S.J., 2004. Venture capital financing, strategic alliances, and the initial public offerings of Internet startups. Journal of Business Venturing. https://doi.org/10.1016/j.jbusvent.2003.03.002
Choi, J., Menon, A., Tabakovic, H., 2021. Using machine learning to revisit the diversification–performance relationship. Strategic Management Journal. https://doi.org/10.1002/smj.3317
Christensen, C.M., 1997. The innovator's dilemma, Management of Innovation and Change. Harvard Business Review Press, Boston, MA.
Cohen, W.M., Levinthal, D.A., 1990. Absorptive capacity: A new perspective on learning and innovation. Adm. Sci. Q. 35, 128.
Cohen, W.M., Nelson, R.R., Walsh, J.P., 2000. Protecting Their Intellectual Assets: Appropriability Conditions and Why U.S. Manufacturing Firms Patent (or Not). Working Paper Series. https://doi.org/10.3386/w7552

Cornelissen, J., Cholakova, M., 2021. Profits Uber everything? The gig economy and the morality of category work. Strategic Organization 19, 722–731.
Devlin, J., Chang, M.-W., Lee, K., Toutanova, K., 2018. BERT: Pre-training of Deep Bidirectional Transformers for Language Understanding. arXiv [cs.CL].
Ethiraj, S.K., Levinthal, D., 2004. Modularity and Innovation in Complex Systems. Manage. Sci. 50, 159–173.
Fleming, L., 2001. Recombinant Uncertainty in Technological Search. Manage. Sci. 47, 117–132.
Fleming, L., Sorenson, O., 2001. Technology as a complex adaptive system: evidence from patent data. Res. Policy 30, 1019–1039.
Galbraith, J.R., 1973. Designing complex organizations, Addison-Wesley series on organization development. Addison Wesley, London, England.
Gans, J., Scott, E.L., Stern, S., 2022. Entrepreneurial strategy: a choice-based approach to entrepreneurship education, in: Annals of Entrepreneurship Education and Pedagogy – 2023. Edward Elgar Publishing, pp. 393–400.
Gans, J.S., Stern, S., Wu, J., 2019. Foundations of entrepreneurial strategy. Strategic Manage. J. 40, 736–756.
Garg, N., Schiebinger, L., Jurafsky, D., Zou, J., 2017. Word Embeddings Quantify 100 Years of Gender and Ethnic Stereotypes. arXiv [cs.CL].
Goncalo, J.A., Chatman, J., Duguid, M., Kennedy, J., 2015. Creativity from Constraint? How Political Correctness Influences Creativity in Mixed-Sex Work Groups.
Guzman, J., Li, A., n.d. Measuring Founding Strategy. https://doi.org/10.31235/osf.io/7cvge
Hamilton, W.L., Leskovec, J., Jurafsky, D., 2016. Diachronic Word Embeddings Reveal Statistical Laws of Semantic Change. arXiv [cs.CL].
Hannan, M.T., Freeman, J., 1984. Structural Inertia and Organizational Change. Am. Sociol. Rev. 49, 149–164.
Hannan, M.T., Freeman, J., 1977. The Population Ecology of Organizations. Am. J. Sociol. 82, 929–964.
Henderson, R., 2021. Innovation in the 21st Century: Architectural Change, Purpose, and the Challenges of Our Time. Manage. Sci. 67, 5479–5488.
Henderson, R., Cockburn, I., 1994. Measuring competence? Exploring firm effects in pharmaceutical research. Strategic Manage. J. 15, 63–84.
Henderson, R.M., Clark, K.B., 1990. Architectural Innovation: The Reconfiguration of Existing Product Technologies and the Failure of Established Firms. Adm. Sci. Q. 35, 9–30.
Hidalgo, C.A., Klinger, B., Barabási, A.-L., Hausmann, R., 2007. The product space conditions the development of nations. Science 317, 482–487.
Hsu, G., Hannan, M.T., 2005. Identities, genres, and organizational forms. Organ. Sci. 16, 474–490.
Jones, B.F., Summers, L.H., 2020. A Calculation of the Social Returns to Innovation (No. w27863). National Bureau of Economic Research. https://doi.org/10.3386/w27863
Joulin, A., Grave, E., Bojanowski, P., Mikolov, T., 2016. Bag of Tricks for Efficient Text Classification. arXiv [cs.CL].
Kaplan, S.N., Sensoy, B.A., Strömberg, P., 2009. Should Investors Bet on the Jockey or the Horse? Evidence from the Evolution of Firms from Early Business Plans to Public Companies. J. Finance 64, 75–115.
Kaplan, S., Vakili, K., 2015. The double-edged sword of recombination in breakthrough innovation. Strategic Manage. J. 36, 1435–1457.
Katila, R., Ahuja, G., 2002. Something old, something new: A longitudinal study of search behavior and new product introduction. Acad. Manage. J. 45, 1183–1194.
Kauffman, S.A., 2002. Investigations. Oxford University Press.
Kauffman, S., Lobo, J., Macready, W.G., 2000. Optimal search on a technology landscape. J. Econ. Behav. Organ. 43, 141–166.
Keyhani, M., Deutsch, Y., Madhok, A., Lévesque, M., 2022. Exploration-exploitation and acquisition likelihood in new ventures. Small Business Economics. https://doi.org/10.1007/s11187-021-00452-1
Kiss, A.N., Barr, P.S., 2015. New venture strategic adaptation: The interplay of belief structures and industry context. Strategic Management Journal. https://doi.org/10.1002/smj.2285
Kozlowski, A.C., Evans, J., 2024. Simulating subjects: The promise and peril of AI stand-ins for social agents and interactions. SocArXiv. https://doi.org/10.31235/osf.io/vp3j2
Kozlowski, A.C., Taddy, M., Evans, J.A., 2019. The Geometry of Culture: Analyzing the Meanings of Class through Word Embeddings. Am. Sociol. Rev. 84, 905–949.
Kulkarni, V., Al-Rfou, R., Perozzi, B., Skiena, S., 2015. Statistically Significant Detection of Linguistic Change, in: Proceedings of the 24th International Conference on World Wide Web, WWW '15. International World Wide Web Conferences Steering Committee, Republic and Canton of Geneva, CHE, pp. 625–635.
Latour, B., 1993. The Pasteurization of France. Harvard University Press.
Leonard-Barton, D., 1993. Core capabilities and core rigidities: A paradox in managing new product development. Long Range Plann. 26, 154.
Levinthal, D.A., 1997. Adaptation on Rugged Landscapes. Manage. Sci. 43, 934–950.
Levy, Goldberg, n.d. Neural word embedding as implicit matrix factorization. Adv. Neural Inf. Process. Syst.
Levy, O., Goldberg, Y., Dagan, I., 2015. Improving Distributional Similarity with Lessons Learned from Word Embeddings. Transactions of the Association for Computational Linguistics 3, 211–225.
Liu, F., Zheng, L., Zheng, J., 2020. HieNN-DWE: A hierarchical neural network with dynamic word embeddings for document level sentiment classification. Neurocomputing. https://doi.org/10.1016/j.neucom.2020.04.084
Ljungqvist, A., Hochberg, Y.V., 2009. Whom You Know Matters: Venture Capital Networks and Investment Performance. SSRN.
Ludwig, J., Mullainathan, S., Rambachan, A., 2024. Large language models: An applied econometric framework. arXiv [econ.EM].
March, J.G., 1991. Exploration and Exploitation in Organizational Learning. Organization Science 2, 71–87.
Marx, M., Gans, J.S., Hsu, D.H., 2014. Dynamic Commercialization Strategies for Disruptive Technologies: Evidence from the Speech Recognition Industry. Manage. Sci. 60, 3103–3123.
McDonald, R.M., Eisenhardt, K.M., 2020. Parallel Play: Startups, Nascent Markets, and Effective Business-model Design. Adm. Sci. Q. 65, 483–523.
McKnight, B., Zietsma, C., 2018. Finding the threshold: A configurational approach to optimal distinctiveness. J. Bus. Venturing 33, 493–512.
Mealy, P., Farmer, J.D., Teytelboym, A., 2019. Interpreting economic complexity. Sci Adv 5, eaau1705.

Mikolov, T., Chen, K., Corrado, G., Dean, J., 2013a. Efficient Estimation of Word Representations in Vector Space. arXiv [cs.CL].
Mikolov, T., Chen, K., Corrado, G., Dean, J., Sutskever, L., Zweig, G., 2013b. word2vec. URL https://code. google. com/p/word2vec 22.
Mozumder, S.I., Rutherford, M.J., Lambert, P.C., 2017. stpm2cr: A flexible parametric competing risks model using a direct likelihood approach for the cause-specific cumulative incidence function. Stata J. 17, 462–489.
Navis, C., Glynn, M.A., 2010. How New Market categories emerge: Temporal dynamics of legitimacy, identity, and entrepreneurship in satellite radio, 1990–2005. Adm. Sci. Q. 55, 439–471.
O'Quigley, J., Xu, R., Stare, J., 2005. Explained randomness in proportional hazards models. Statistics in Medicine. https://doi.org/10.1002/sim.1946
Page, S.E., 2019. The Diversity Bonus: How Great Teams Pay Off in the Knowledge Economy. Princeton University Press.
Page, S.E., 2008. The Difference: How the Power of Diversity Creates Better Groups, Firms, Schools, and Societies - New Edition. Princeton University Press.
Pennington, J., Socher, R., Manning, C., 2014. Glove: Global vectors for word representation, in: Proceedings of the 2014 Conference on Empirical Methods in Natural Language Processing (EMNLP). pp. 1532–1543.
Ries, E., 2011. The Lean Startup: How Today's Entrepreneurs Use Continuous Innovation to Create Radically Successful Businesses. Crown.
Rivkin, J.W., Siggelkow, N., 2003. Balancing search and stability: Interdependencies among elements of organizational design. Manage. Sci. 49, 290–311.
Roche, M.P., Conti, A., Rothaermel, F.T., 2020. Different founders, different venture outcomes: A comparative analysis of academic and non-academic startups. Research Policy. https://doi.org/10.1016/j.respol.2020.104062
Rosenbloom, R.S., Christensen, C.M., 1994. Technological discontinuties, organizational capabilities, and strategic commitments. Ind. Corp. Chang. 3, 655–685.
Royston, P., 2006. Explained Variation for Survival Models. The Stata Journal: Promoting communications on statistics and Stata. https://doi.org/10.1177/1536867x0600600105
Sanchez, R., Mahoney, J.T., 1996. Modularity, flexibility, and knowledge management in product and organization design. Strategic Manage. J. 17, 63–76.
Sarasvathy, S.D., 2001. Causation and Effectuation: Toward a Theoretical Shift from Economic Inevitability to Entrepreneurial Contingency. AMRO 26, 243–263.
Savin, I., Chukavina, K., Pushkarev, A., 2022. Topic-based classification and identification of global trends for startup companies. Small Business Economics. https://doi.org/10.1007/s11187-022-00609-6
Shane, S., 2000. Prior knowledge and the discovery of entrepreneurial opportunities. Organ. Sci. 11, 448–469.
Shiller, R.J., Brown, B., 2019. NARRATIVE ECONOMICS: HOW STORIES GO VIRAL AND DRIVE MAJOR ECONOMIC EVENTS. The Quarterly Journal of Austrian Economics 22, 620+.
Shi, Y., Sorenson, O., Waguespack, D.M., 2017. Temporal Issues in Replication: The Stability of Centrality-Based Advantage. Sociological Science 4, 107–122.
Simon, H.A., 1991. The Architecture of Complexity, in: Klir, G.J. (Ed.), Facets of Systems Science. Springer US, Boston, MA, pp. 457–476.

Sorenson, O., 2018. Social networks and the geography of entrepreneurship. Small Bus. Econ. 51, 527–537.
Stark, D., 2011. The Sense of Dissonance: Accounts of Worth in Economic Life. Princeton University Press, Princeton, NJ.
Taylor, A., Greve, H.R., 2006. Superman or the Fantastic Four? knowledge combination And experience in Innovative Teams. AMJ 49, 723–740.
Ter Wal, A.L.J., Alexy, O., Block, J., Sandner, P.G., 2016. The Best of Both Worlds: The Benefits of Open-specialized and Closed-diverse Syndication Networks for New Ventures' Success. Adm. Sci. Q. 61, 393–432.
Thompson, J.D., 1967. Organizations in action. McGraw-Hill, New York, NY.
Tseng, M.M., Wang, Y., Jiao, R.J., 2019. Modular Design, in: CIRP Encyclopedia of Production Engineering. Springer, Berlin, Heidelberg, pp. 1226–1235.
Tushman, M.L., Anderson, P., 1986. Technological discontinuities and organizational environments. Adm. Sci. Q. 31, 439.
van Binsbergen, J.H., Bryzgalova, S., Mukhopadhyay, M., Sharma, V., n.d. (Almost) 200 Years of News-Based Economic Sentiment. SSRN Electronic Journal. https://doi.org/10.2139/ssrn.4261249
Wagner, A., Ortman, S., Maxfield, R., 2016. From the primordial soup to self-driving cars: standards and their role in natural and technological innovation. J. R. Soc. Interface 13, 20151086.
Wimsatt, W.C., 2014. Entrenchment and scaffolding: an architecture for a theory of cultural change. Developing scaffolds in evolution, culture, and cognition 17, 77–105.
Yao, Z., Sun, Y., Ding, W., Rao, N., Xiong, H., 2018. Dynamic Word Embeddings for Evolving Semantic Discovery, in: Proceedings of the Eleventh ACM International Conference on Web Search and Data Mining, WSDM '18. Association for Computing Machinery, New York, NY, USA, pp. 673–681.
Youn, H., Strumsky, D., Bettencourt, L.M.A., Lobo, J., 2015. Invention as a combinatorial process: evidence from US patents. J. R. Soc. Interface 12, 20150272.
Zhang, Y., Jatowt, A., Bhowmick, S.S., Tanaka, K., 2016. The Past is Not a Foreign Country: Detecting Semantically Similar Terms across Time. IEEE Trans. Knowl. Data Eng. 28, 2793–2807.
Zuckerman, E.W., 1999. The Categorical Imperative: Securities Analysts and the Illegitimacy Discount. Am. J. Sociol. 104, 1398–1438.

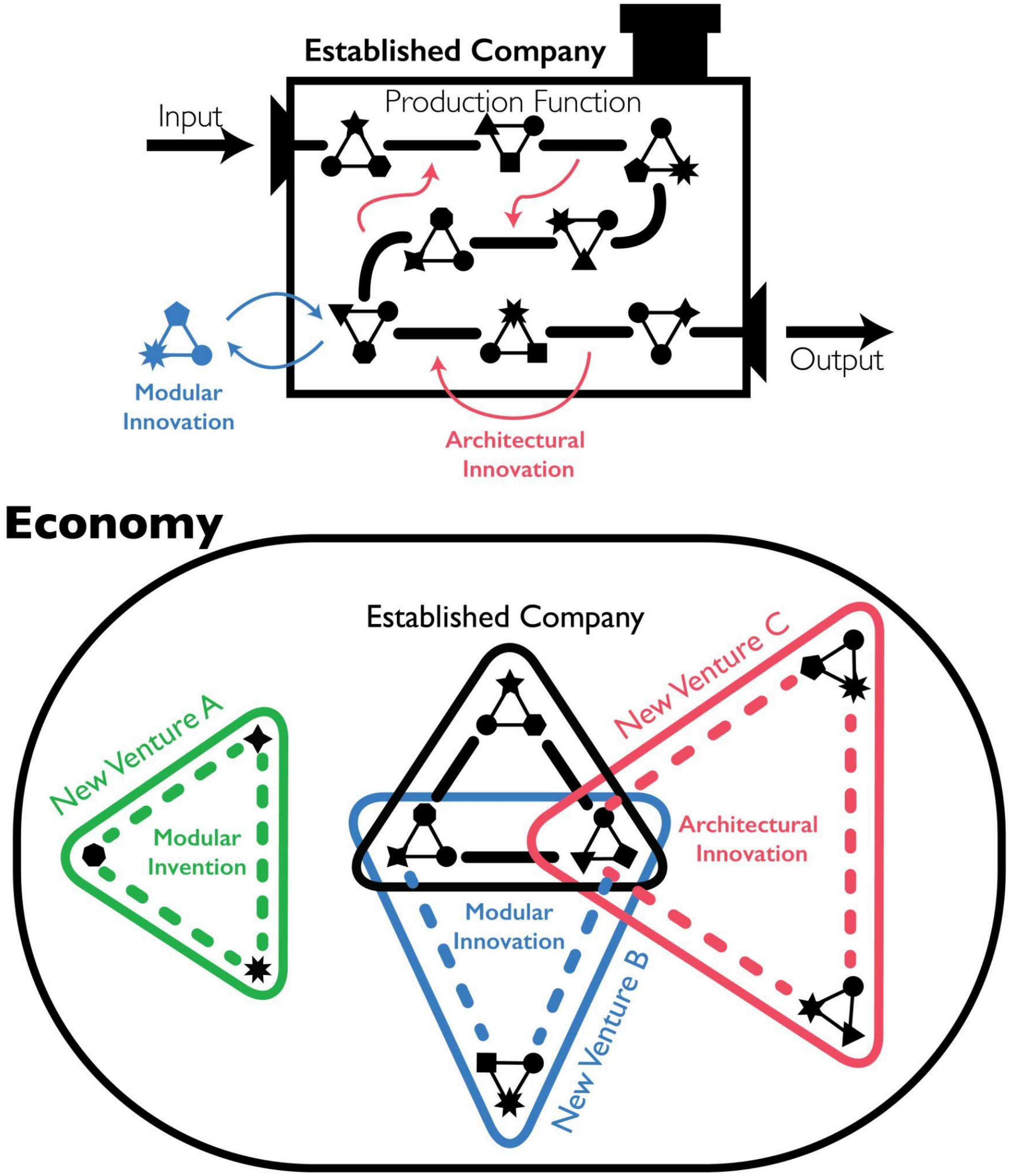

**Figure 1**. Conceptual Figure of the Three Innovation Strategies

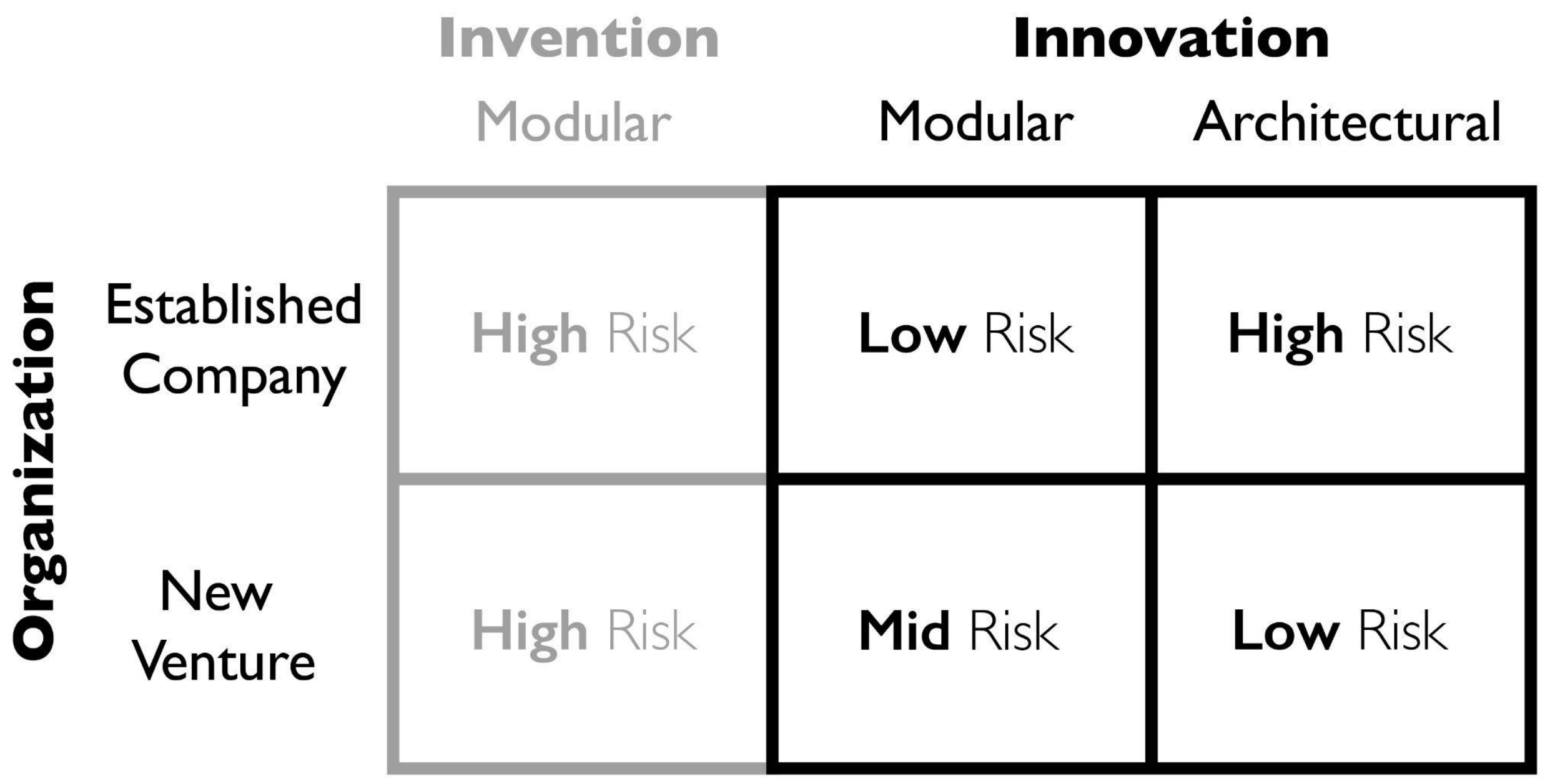

**Figure 2. Theoretical Framework for Entrepreneurial Innovation Strategies**

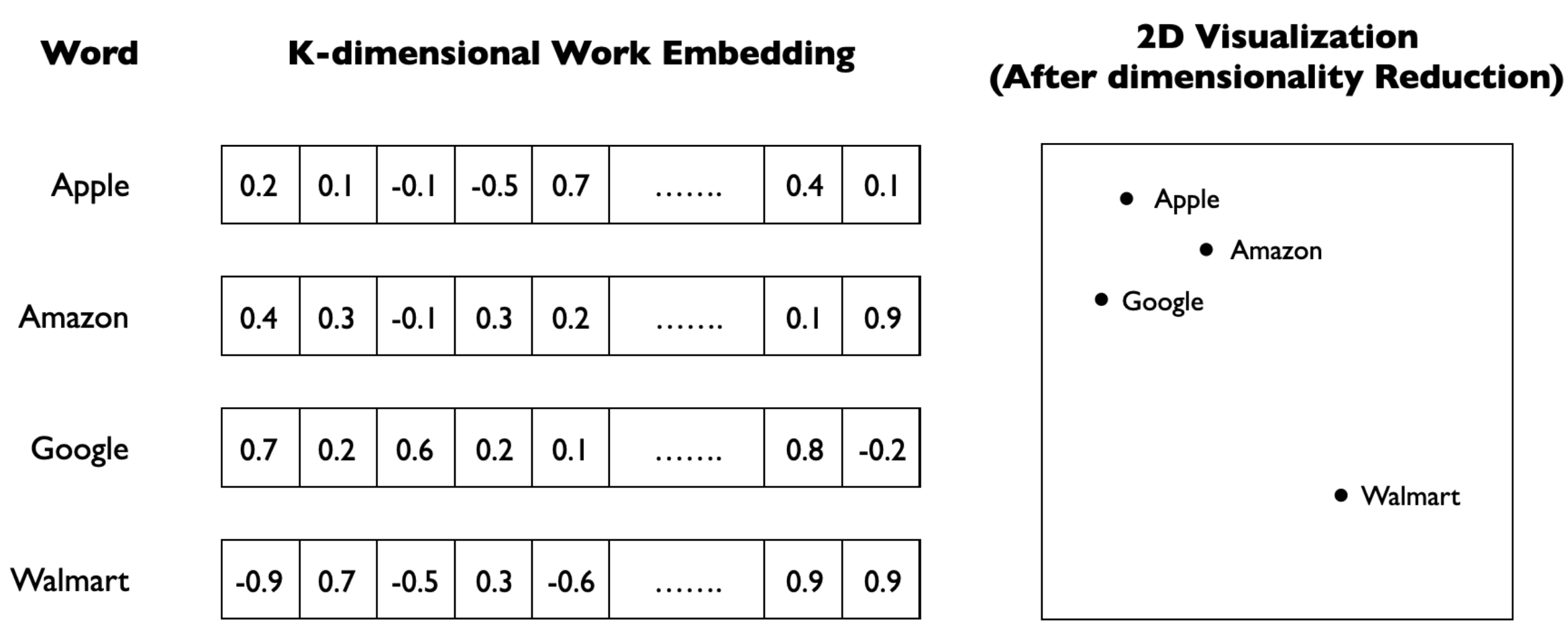

**Figure 3. Conceptual Figure representing Word Vectors in an Embedding Space.**

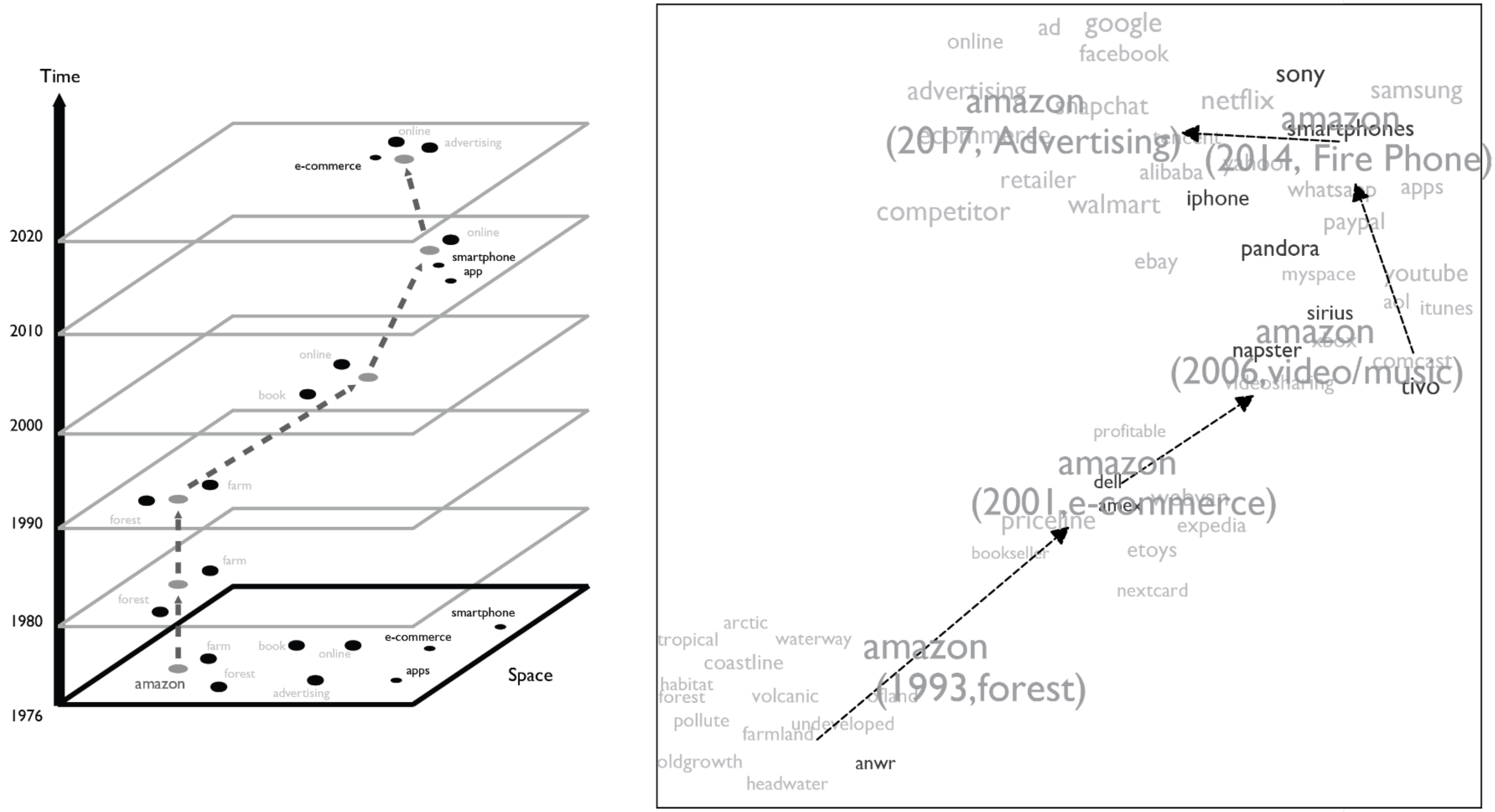

**Figure 4. Dynamic Business Landscape.***

*Left, a conceptual representation of the layered construction of our semantic space representing business discourse. Right, an empirical figure, orthogonal to the one on the left, which flattens the overlapping layers, highlighting the movement of Amazon through conceptual space over time and its neighborhoods of most similar concepts in 1993, 2001, 2006, 2014, and 2017.

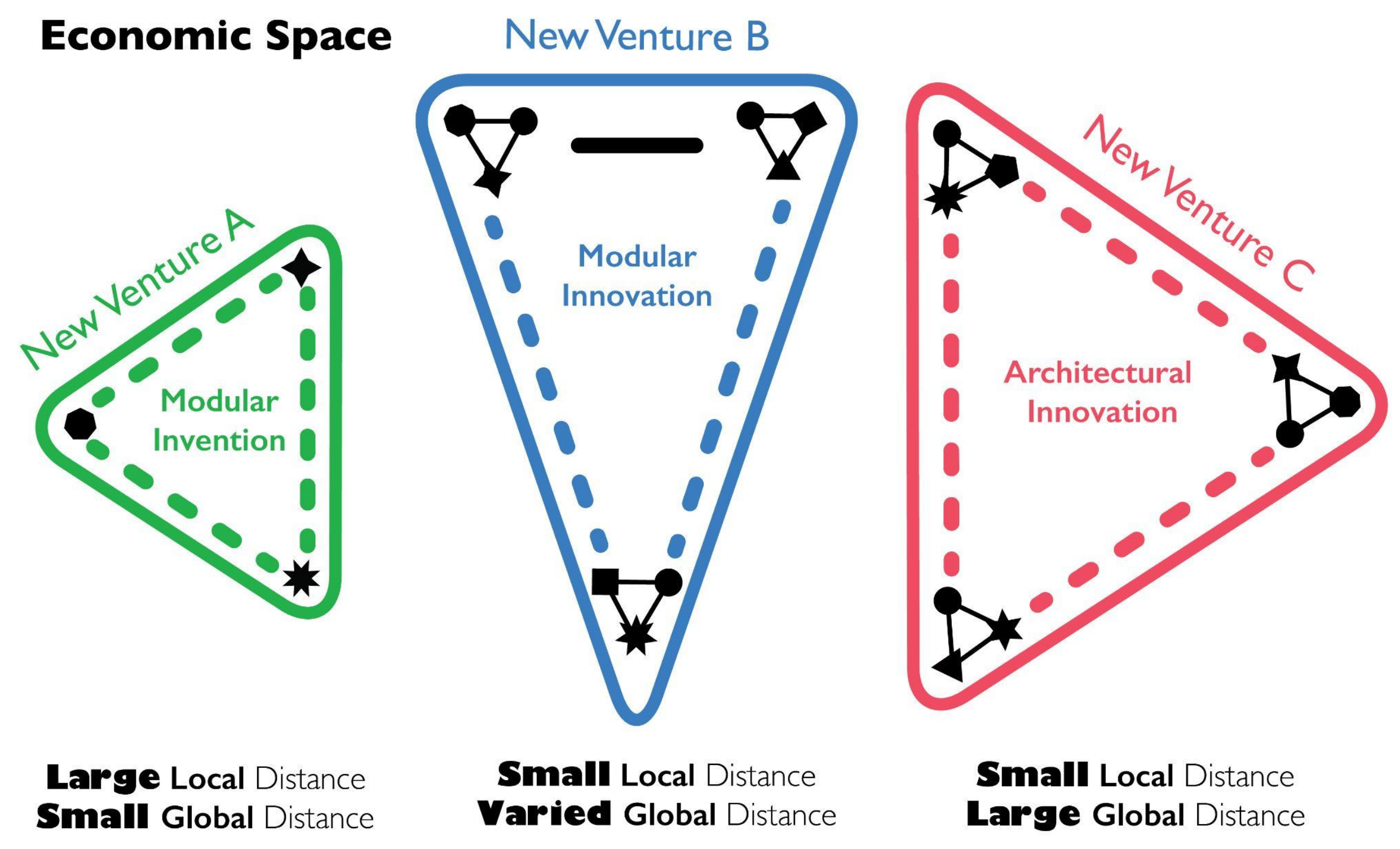

**Figure 5. Representation of Meaures in Embedding Space.**

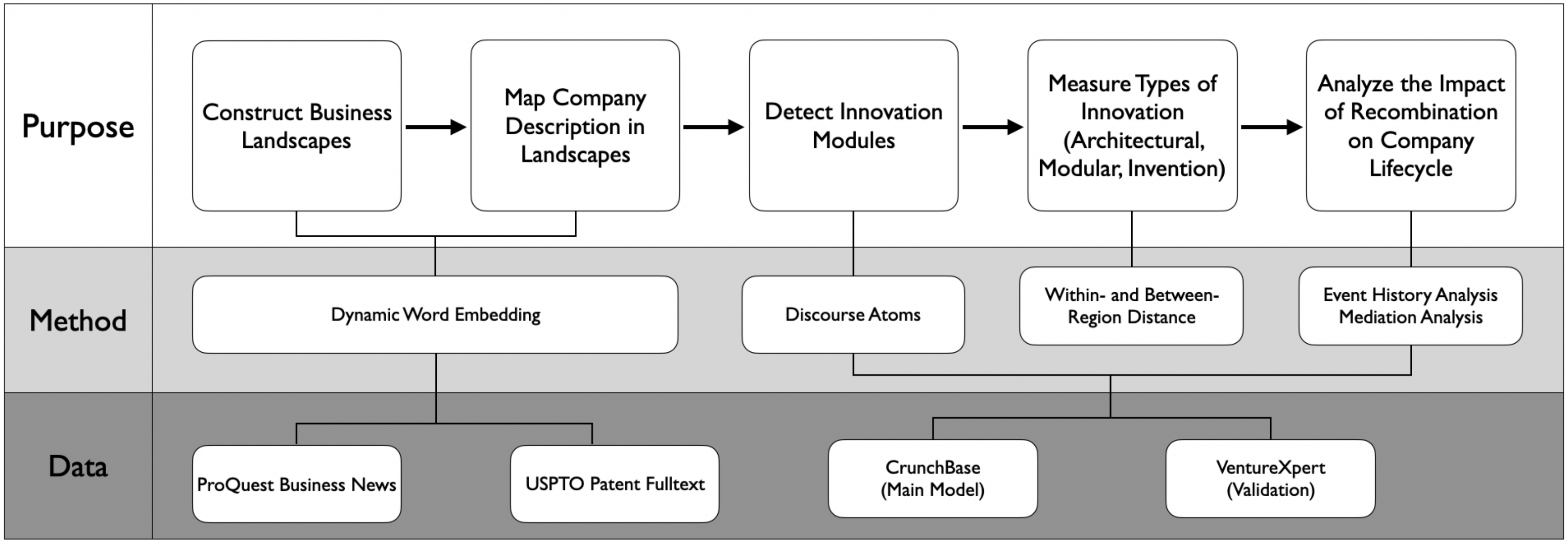

**Figure 6. Data Analysis Process.**

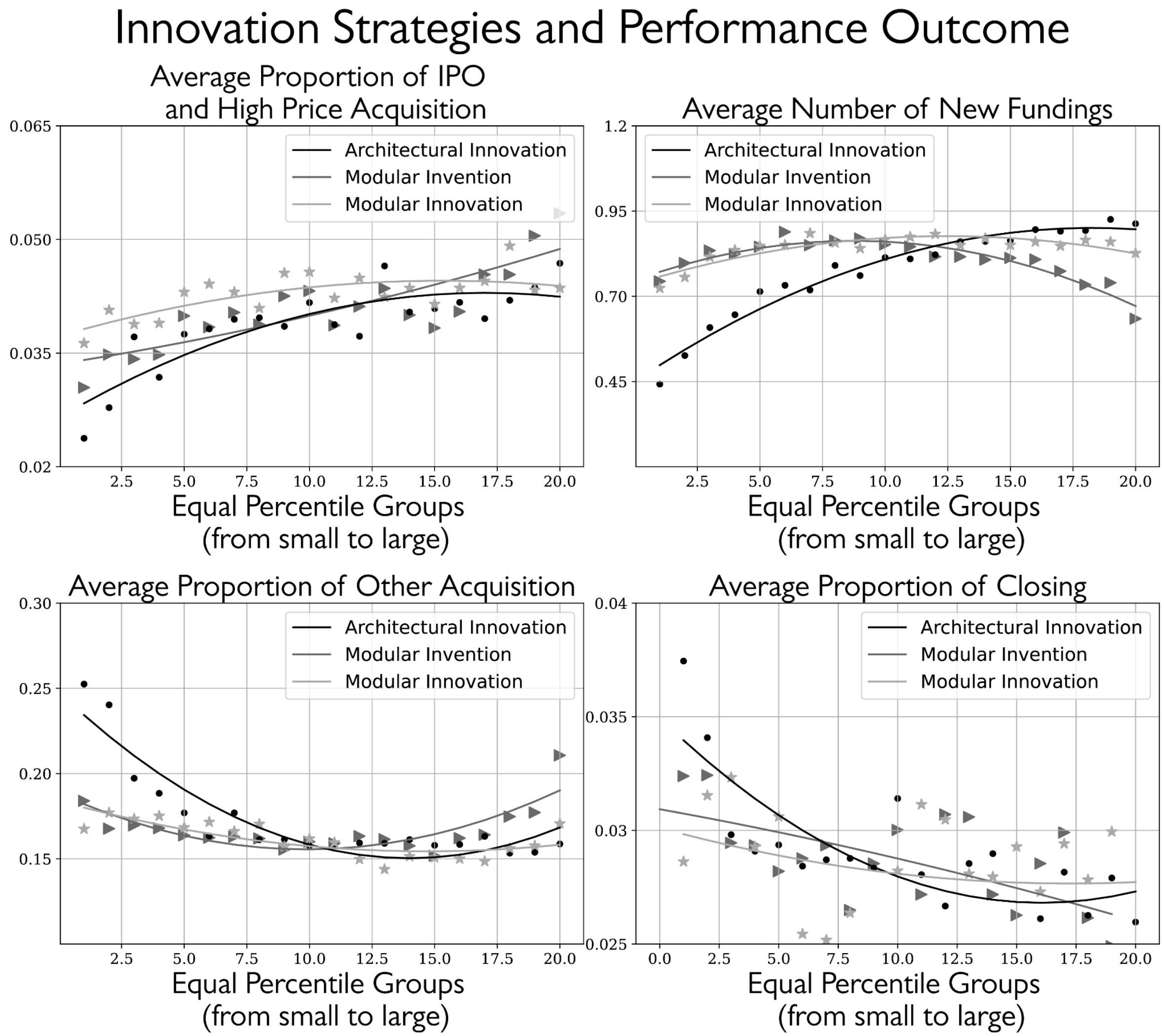

**Figure 7. Descriptive relationship between Independent Variables and Venture Outcomes.**

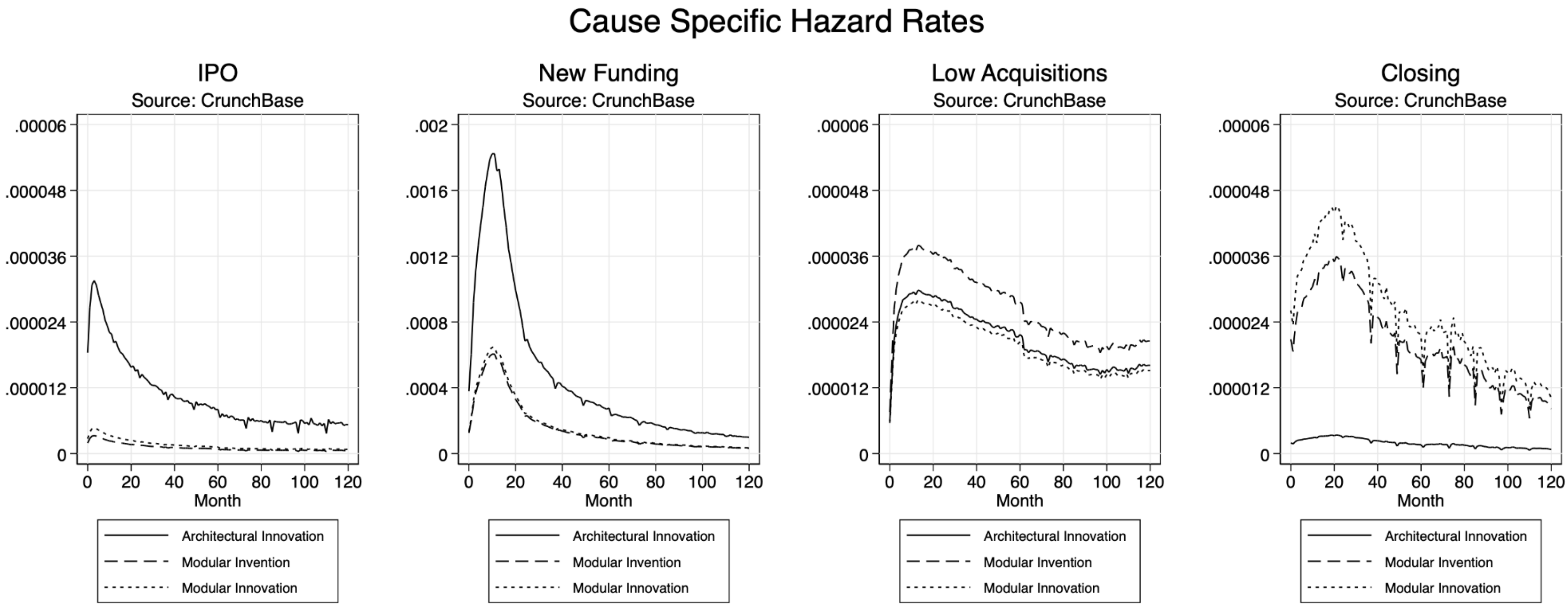

**Figure 8. Hazard Rate Prediction for "Ideal" Companies.**

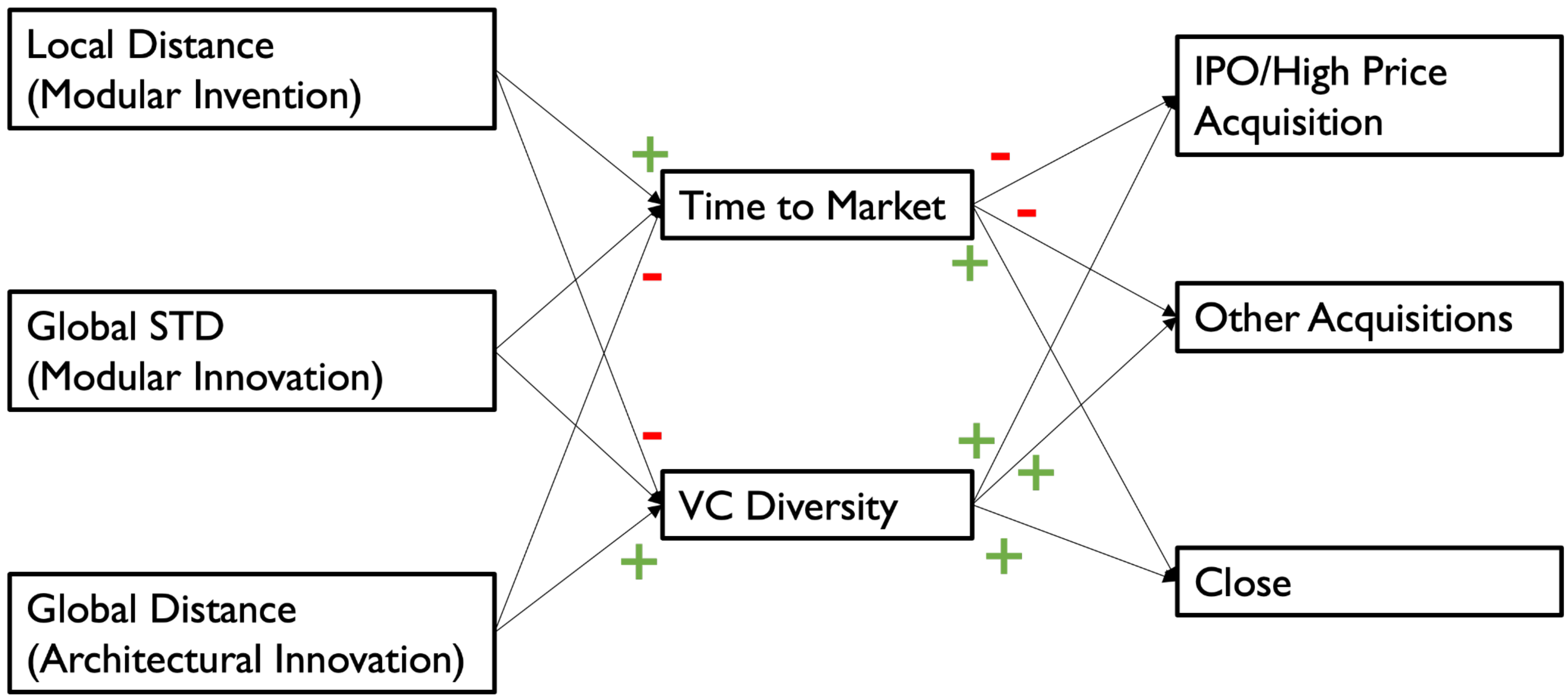

**Figure 9. Summary of Indirect Path Effects based on Generalized Structural Equation Modeling.**

Table 1. Examples for Discourse Atoms (1990 and 2010)

| Example Discourse Atoms | Techno-Social Elements in the Atom |
|---|---|
| **1990 Examples** | |
| Atom 1 | gasification, petroleum, hydropower, transportation, manufacturing, construction |
| Atom 2 | cabletv, telegraph, widescreen, polaroid, publishing, broadcasting, roadshow |
| Atom 3 | androgenic, cerebrum, psychographics, dating, boardroom, pawnbroker |
| Atom 4 | archaeology, plato, folklore, mythology, spiritualism, melodrama, filmmaker |
| Atom 5 | thermocouple, disk, pcb, tomography, apparatus, probe, channel, connection |
| Atom 6 | photocurrents, entropy, spectroradiometer, measuring, meter, transition, sense |
| Atom 7 | gemstone, linen, onyx, maraschino, lace, towel, womenswear, pottery, style |
| Atom 8 | biofuel, petrol, uranium, biogas, refine, mineral, papermaking, automaking |
| Atom 9 | cephalosporin, arthroscopy, aspirin, surgery, anesthesia, prevention |
| Atom 10 | hyperboloid, ellipsoidal, optic, vertices, eyelens, intermetal, viewfinder, zigzag |
| **2010 Examples** | |
| Atom 1 | pantone, herringbone, textile, fashion, designer, shoe, hat, wear |
| Atom 2 | pycnometer, biofluid, microbiocide, circulation, recycling, fuel, injection |
| Atom 3 | smartphone, cryptocurrency, webcam, chat, video, networking, advertising |
| Atom 4 | ultrasonication,deaggregation, mineralogy, cleanness, crystallizer, fiberizing |
| Atom 5 | overmolded, laminate, caulking, casting, glass, plastic, pressboard |
| Atom 6 | sibutramine, isocarboxazid, linezolid, therapy, treatment, vaccine, preventative |
| Atom 7 | riskmetrics, telebank, microfin, megadeals, coinvest, autofinance |
| Atom 8 | circuitry, multiplexer, synchronous, converter, generator, signal, terminal |
| Atom 9 | lans, uplink, vocoding, websockets, networks, brokering, communications |
| Atom 10 | histologic, vascularity, tracheotomy, neogenesis, surgery, anesthesia |

Table 2. Descriptive Statistics of Variables

| | Crunchbase (n=197978) | | VentureXpert (n=63018) | |
|---|---|---|---|---|
| | mean | std | mean | std |
| ***Dependent Variables*** | | | | |
| Average number of accepted funding to date | 0.758 | 1.471 | 2.455 | 2.626 |
| Proportion of IPO or high price acquisition | 0.038 | 0.202 | 0.099 | 0.300 |
| Proportion of middle or low-price acquisition | 0.173 | 0.378 | 0.172 | 0.377 |
| Proportion of closing | 0.029 | 0.168 | 0.130 | 0.336 |
| ***Independent Variables*** | | | | |
| Local Distance (Modular Invention) | 0.140 | 0.106 | 0.128 | 0.092 |
| Global STD (Modular Innovation) | 0.141 | 0.035 | 0.134 | 0.033 |
| Global Distance (Architectural Innovation) | 0.449 | 0.058 | 0.450 | 0.065 |
| ***Controls*** | | | | |
| Element Familiarity | 2.399 | 1.426 | 1.067 | 0.680 |
| Dummy: Usage of social media | 0.755 | 0.430 | Not Applicable | |
| Dummy: founded by women or minorities | 0.123 | 0.346 | Not Applicable | |
| Market growth rate in the year of establishment | -0.068 | 0.228 | Not Applicable | |
| Number of valid elements | 30.702 | 20.564 | 42.362 | 22.468 |
| Dummy: no technical elements at all | 0.189 | 0.392 | 0.164 | 0.370 |
| Uncommon elements | 0.021 | 0.143 | 0.011 | 0.103 |

Table 3. Correlation Table for the Major Variables

| | (1) | (2) | (3) | (4) | (5) | (6) | (7) | (8) | (9) | (10) |
|---|---|---|---|---|---|---|---|---|---|---|
| (1) Local Distance (Modular Invention) | 1.000 | | | | | | | | | |
| (2) Global STD (Modular Innovation) | 0.006*** | 1.000 | | | | | | | | |
| | (0.000) | | | | | | | | | |
| (3) Global Distance (Architectural Innovation) | -0.311*** | 0.408*** | 1.000 | | | | | | | |
| | (0.000) | (0.000) | | | | | | | | |
| (4) Element familiarity | 0.022*** | -0.112*** | -0.231*** | 1.000 | | | | | | |
| | (0.000) | (0.000) | (0.000) | | | | | | | |
| (5) Dummy: usage of social media | -0.077*** | -0.017*** | 0.063*** | 0.144*** | 1.000 | | | | | |
| | (0.000) | (0.000) | (0.000) | (0.000) | | | | | | |
| (6) Dummy: founded by women or minorities | -0.030*** | -0.025*** | 0.014*** | 0.087*** | 0.109*** | 1.000 | | | | |
| | (0.000) | (0.000) | (0.000) | (0.000) | (0.000) | | | | | |
| (7) Market growth rate | 0.010*** | 0.060*** | 0.052*** | -0.231*** | -0.053*** | -0.118*** | 1.000 | | | |
| | (0.000) | (0.000) | (0.000) | (0.000) | (0.000) | (0.000) | | | | |
| (8) Number of valid elements | -0.283*** | 0.071*** | 0.449*** | -0.135*** | 0.093*** | 0.009*** | 0.043*** | 1.000 | | |
| | (0.000) | (0.000) | (0.000) | (0.000) | (0.000) | (0.000) | (0.000) | | | |
| (9) Dummy: no technical elements at all | 0.232*** | -0.220*** | -0.437*** | 0.152*** | -0.020*** | -0.003* | -0.004** | -0.335*** | 1.000 | |
| | (0.000) | (0.000) | (0.000) | (0.000) | (0.000) | (0.100) | (0.017) | (0.000) | | |
| (10) Uncommon elements | 0.041*** | -0.063*** | -0.102*** | -0.018*** | -0.025*** | -0.027*** | 0.027*** | -0.069*** | -0.071*** | 1.000 |

| | | | | | | | | | |
|---|---|---|---|---|---|---|---|---|---|
| | (0.000) | (0.000) | (0.000) | (0.000) | (0.000) | (0.000) | (0.000) | (0.000) | (0.000) |

*Note: *p < .05, **p < .01, ***p < .001, two-tailed test.*

Table 4. FPM Model for Crunchbase Data

| | **Model 1: the Control Model** | | | | **Model 2: the Full Model** | | | |
|---|---|---|---|---|---|---|---|---|
| | IPO/High Price Acquisition | New Funding | Other Acquisition | Close | IPO/High Price Acquisition | New Funding | Other Acquisition | Close |
| Local Distance (Modular Invention) | | | | | -0.939*** | -0.424*** | 0.517*** | 0.157 |
| | | | | | (0.146) | (0.029) | (0.043) | (0.123) |
| Global STD (Modular Innovation) | | | | | -0.603 | -0.952*** | -0.474** | 2.170*** |
| | | | | | (0.387) | (0.083) | (0.152) | (0.387) |
| Global Distance (Architectural Innovation) | | | | | 2.258*** | 1.121*** | -0.004 | -2.837*** |
| | | | | | (0.299) | (0.062) | (0.103) | (0.261) |
| Elements Familiarity | -0.176$^{***}$ | -0.001 | 0.086$^{***}$ | -0.019$^{*}$ | -0.139$^{***}$ | 0.027$^{***}$ | 0.062$^{***}$ | -0.067$^{***}$ |
| | (0.015) | (0.002) | (0.003) | (0.009) | (0.016) | (0.002) | (0.004) | (0.010) |
| Dummy: Usage of social media | -0.020 | 0.192$^{***}$ | -0.448$^{***}$ | -0.442$^{***}$ | -0.059$^{*}$ | 0.174$^{***}$ | -0.403$^{***}$ | -0.397$^{***}$ |
| | (0.027) | (0.006) | (0.012) | (0.030) | (0.026) | (0.006) | (0.012) | (0.030) |
| Dummy: founded by women or minorities | -0.428$^{***}$ | 0.303$^{***}$ | -0.413$^{***}$ | -0.279$^{***}$ | -0.558$^{***}$ | 0.292$^{***}$ | -0.443$^{***}$ | -0.267$^{***}$ |
| | (0.060) | (0.007) | (0.027) | (0.044) | (0.055) | (0.007) | (0.026) | (0.044) |
| Market growth rate in the year of establishment | 0.013 | 0.033$^{*}$ | 0.041 | -0.168 | -0.037 | 0.012 | 0.051 | -0.133 |
| | (0.066) | (0.015) | (0.030) | (0.088) | (0.066) | (0.015) | (0.031) | (0.088) |
| Text Features (length, missing, unusual words) | YES | YES | YES | YES | YES | YES | YES | YES |
| Year, Location, Industry and Funding History | YES | YES | YES | YES | YES | YES | YES | YES |
| Constant | -4.225$^{***}$ | -4.034$^{***}$ | -4.812$^{***}$ | -7.201$^{***}$ | -5.079$^{***}$ | -4.320$^{***}$ | -4.699$^{***}$ | -6.388$^{***}$ |
| | (0.076) | (0.056) | (0.044) | (0.339) | (0.150) | (0.062) | (0.060) | (0.355) |
| Goodness of Fit | | $\rho_k^2$=0.390, $R^2$=0.280 | | | | $\rho_k^2$=0.399, $R^2$=0.287 | | |

*Note: ***p < .001, **p < .01, *p < .05, two-tailed test.*

Table 5. Regressing Meditators on Main Independent Variables

| | Samples with Seed Round Records (N=49304) | | Samples with both Seed Round and Early Round Records (N=19979) | |
|---|---|---|---|---|
| | Time to Market | VC Diversity | Time to Market | VC Diversity |
| Local Distance (Modular Invention) | 13.188*** | -0.104*** | -0.952 | -0.127*** |
| | (2.327) | (0.019) | (1.693) | (0.038) |
| Global STD (Modular Innovation) | 8.582 | -0.011 | -2.401 | -0.021 |
| | (6.049) | (0.050) | (4.058) | (0.091) |
| Global Distance (Architectural Innovation) | -34.570*** | 0.135*** | -8.498** | 0.140* |
| | (4.246) | (0.035) | (2.965) | (0.066) |
| Technology Elements Familiarity | 1.488** | 0.000 | 1.581*** | 0.027** |
| | (0.541) | (0.005) | (0.445) | (0.010) |
| Dummy: Usage of social media | 1.351** | 0.025*** | 1.131*** | -0.006 |
| | (0.468) | (0.004) | (0.292) | (0.007) |
| Dummy: founded by women or minorities | 1.704*** | 0.035*** | 1.781*** | 0.029*** |
| | (0.464) | (0.004) | (0.316) | (0.007) |
| Market growth rate in the year of establishment | 26.883*** | -0.018 | -1.934** | -0.029 |
| | (1.138) | (0.009) | (0.733) | (0.016) |
| Text Features | YES | YES | YES | YES |
| Year, Location, Industry, Funding History | YES | YES | YES | YES |
| Constant | 43.642*** | 0.099** | 16.464*** | 0.392*** |
| | (4.320) | (0.036) | (3.568) | (0.080) |
| N | 49269 | 49269 | 19976 | 19976 |
| Goodness of Fit (Adjusted R2) | 0.212 | 0.148 | 0.183 | 0.073 |

*Note: ***p < .001, **p < .01, *p < .05, two-tailed test.*

Table 6. GSEM Model for Mediation Analysis

| | IPO and High Price Acquisition | Other Acquisition | Close | Time to Market | VC Diversity for Early Rounds |
|---|---|---|---|---|---|
| Time to Market | -0.009*** | -0.003*** | 0.005*** | | |
| | (0.001) | (0.000) | (0.000) | | |
| VC Diversity for Early Rounds | 0.614*** | 0.953*** | 0.191** | | |
| | (0.085) | (0.040) | (0.062) | | |
| Local Distance (Modular Invention) | -1.668** | 0.199 | 0.359 | 13.217*** | -0.104*** |
| | (0.555) | (0.192) | (0.226) | (2.326) | (0.019) |
| Global STD (Modular Innovation) | -2.777* | -1.390** | 1.567** | 8.832 | -0.010 |
| | (1.137) | (0.489) | (0.590) | (6.046) | (0.050) |
| Global Distance | 3.932*** | 0.800* | -2.091*** | -34.658*** | 0.134*** |
| (Architectural Innovation) | (0.858) | (0.352) | (0.402) | (4.245) | (0.035) |
| Other Controls | YES | YES | YES | YES | YES |
| Constant | -1.875*** | -3.404*** | 0.570 | 43.804*** | 0.099** |
| | (0.525) | (0.597) | (0.369) | (4.318) | (0.036) |

*Note: ***p < .001, **p < .01, *p < .05, two-tailed test. Control variables are the same as in FPM. Generalized structural equation model using bootstrapping (1,000 repetitions), bias-corrected coefficients, and robust standard errors.*

Table 7. Path Effects for GSEM

| | **Direct Effects** | **Indirect Effects: Time to Market** | **Indirect Effects: VC Diversity** | **Total Effects** |
|---|---|---|---|---|
| **DV: IPO and High Price Acquisition** | | | | |
| Local Distance (Modular Invention) | -1.668** | -0.124*** | -0.064*** | -1.856*** |
| | [-2.756, -0.580] | [-0.174, -0.075] | [-0.093, -0.035] | [-2.945, -0.768] |
| Global STD (Modular Innovation) | -2.777* | -0.083 | -0.006 | -2.866* |
| | [-5.005, -0.549] | [-0.196, 0.030] | [-0.067, 0.055] | [-5.098, -0.634] |
| Global Distance (Architectural Innovation) | 3.932*** | 0.326*** | 0.082*** | 4.341*** |
| | [2.251, 5.613] | [0.224, 0.429] | [0.034, 0.130] | [2.657, 6.025] |
| **DV: Other Acquisitions** | | | | |
| Local Distance (Modular Invention) | 0.199 | -0.036*** | –0.099*** | 0.064 |
| | [-0.177, 0.574] | [-0.051, –0.021] | [-0.136, -0.062] | [-0.314, 0.441] |
| Global STD (Modular Innovation) | -1.390** | -0.024 | -0.009 | -1.423** |
| | [-2.349, -0.431] | [-0.057, 0.009] | [-0.103, 0.085] | [-2.388, -0.458] |
| Global Distance (Architectural Innovation) | 0.800* | 0.094*** | 0.128*** | 1.022** |
| | [0.110, 1.490] | [0.062, 0.126] | [0.061, 0.195] | [0.329, 1.716] |
| **DV: Close** | | | | |
| Local Distance (Modular Invention) | 0.359 | 0.061*** | -0.020** | 0.399+ |
| | [-0.084, 0.801] | [0.037, 0.084] | [-0.034, -0.005] | [-0.043, 0.842] |
| Global STD (Modular Innovation) | 1.567** | 0.040 | -0.002 | 1.605** |
| | [0.411, 2.722] | [-0.014, 0.095] | [-0.021, 0.017] | [0.449, 2.762] |
| Global Distance (Architectural Innovation) | -2.091*** | -0.159*** | 0.026* | -2.224*** |
| | [-2.879, -1.303] | [-0.206, -0.112] | [0.005, 0.047] | [-3.012, -1.435] |

*Note: ***p < .001, **p < .01, *p < .05, +p < .1. two-tailed test.*

# Appendix

## Appendix A. The ProQuest Database and Dynamic Business Landscape

### A.1 The ProQuest Database Overview

For business publications, we predominantly draw upon the ProQuest ABI/INFORM Collection, one of the most comprehensive corpora focused on national and international business coverage. In addition to newspapers and magazines, it includes trade journals, reports, economic blogs, scholarly journals, and more. Detailed information about this corpus can be found at ProQuest official website at https://proquest.libguides.com/abiinformcollection

In this study, we retain English newspapers and magazines published under the category of "Business and Economics" in the United States. The catalog of this corpora is listed in

Table A1.

Table A1. Included Business Publications

| TITLE | Pub Type | Publisher |
|---|---|---|
| Aggregates Manager (Online) | Magazines | Randall Reilly |
| American Banker | Newspapers | SourceMedia |
| American Banker (pre-1997 Fulltext) | Newspapers | SourceMedia |
| Army/Navy Store & Outdoor Merchandiser | Magazines | SouthComm Business Media LLC |
| Atlanta Tribune, The | Magazines | Atlanta Tribune: The Magazine |
| Automatic Merchandiser | Magazines | SouthComm Business Media LLC |
| Barron's (1921-1942) | Newspapers | Dow Jones & Company Inc |
| Better Investing | Magazines | National Association of Investment Clubs |
| Bloomberg Businessweek | Magazines | Bloomberg Finance LP |
| Bond Buyer, The | Newspapers | SourceMedia |
| Boston Business Journal, The | Magazines | American City Business Journals |
| Business and Economic Review | Magazines | Moore School of Business |
| Business Insider | Newspapers | Insider, Inc. |
| Business Intelligence Journal | Magazines | Data Warehousing Institute |
| Business Month | Magazines | Goldhirsh Group, Inc. |
| CC News | Newspapers | United Publications, Inc. |
| Channel Executive (Online) | Magazines | Jameson Publishing/Vert Markets |
| Charleston Regional Business Journal | Newspapers | BridgeTower Media Holding Company |
| Coast Business | Magazines | Ship Island Holding Company |
| Columbia Regional Business Report | Newspapers | BridgeTower Media Holding Company |
| Contract Management | Magazines | National Contract Management Association |
| Corporate Adviser (Online) | Magazines | Centaur Media USA Inc. (A member of Centaur Plc Group) |
| Cost Management | Magazines | Thomson Reuters (Tax & Accounting) Inc |
| CPI Detailed Report | Magazines | Superintendent of Documents, Labor Department |
| Crain's Chicago Business | Magazines | Crain Communications, Incorporated |
| Customer Relationship Management | Magazines | Information Today, Inc. |
| Daily Breeze | Newspapers | Los Angeles Newspaper Group |
| Daily Record | Newspapers | BridgeTower Media Holding Company |
| Deal.com, The | Magazines | The Deal LLC |
| Dealer Magazine | Magazines | Emerald Expositions LLC |
| Diablo Business | Magazines | Diablo Business |
| Diversity Factor | Magazines | Diversity Factor |
| Diversity Suppliers & Business Magazine | Magazines | Hispanic Times |
| Dollars & Sense | Magazines | Economic Affairs Bureau |
| Economist, The | Magazines | The Economist Intelligence Unit N.A., Incorporated |
| Employment and Earnings (Online) | Magazines | Superintendent of Documents |
| Fair Employment Practices Guidelines | Magazines | Aspen Publishers, Inc. |
| Fast Company | Magazines | Mansueto Ventures LLC |
| Field Technologies (Online) | Magazines | Jameson Publishing/Vert Markets |
| Finance and Commerce | Newspapers | BridgeTower Media Holding Company |
| Financial Planning | Magazines | SourceMedia |
| Financial World | Magazines | Financial World Partners |
| Florida Trend | Magazines | Trend Magazine, Inc. |
| Forbes Life | Magazines | Forbes |
| Forbes ASAP | Magazines | Forbes |
| Fundweb | Magazines | Centaur Media USA Inc. (A member of Centaur Plc Group) |
| GSA Business Journal | Newspapers | BridgeTower Media Holding Company |
| Harvard Asia Pacific Review | Magazines | Harvard Asia Pacific Review |
| Harvard Business Review | Magazines | Harvard Business Review |
| Inc | Magazines | Mansueto Ventures LLC |
| Industrial Worker | Newspapers | Industrial Workers of the World |
| InfoAmericas | Newspapers | Global Network Content Services LLC, DBA Noticias Financieras LLC |
| Information Management | Magazines | IGI Global |
| Inland Empire Business Journal | Magazines | Daily Planet Publishing Inc. |
| IPS - Inter Press Service | Newspapers | Global Network Content Services LLC, DBA Noticias Financieras LLC |
| Journal of Commerce | Newspapers | IHS Maritime & Trade |
| License! Global | Magazines | MultiMedia Healthcare Inc. |
| Life Science Leader (Online) | Magazines | Jameson Publishing/Vert Markets |
| Louisville | Magazines | Louisville Magazine Inc. |
| Management Accounting (1986-1986) | Magazines | Institute of Management Accountants |
| Management Accounting (pre-1986) | Magazines | Institute of Management Accountants |
| Marketing Week (Online) | Magazines | Centaur Media USA Inc. (A member of Centaur Plc Group) |
| McKinsey Insights | Magazines | McKinsey & Company, Inc. |
| Mecklenburg Times, The | Newspapers | BridgeTower Media Holding Company |
| Monitor Global Outlook | | |
| Minority Business Entrepreneur | Magazines | Full Stride Media |
| Modern Trader | Magazines | Alpha Pages |
| Money Marketing (Online) | Magazines | Centaur Media USA Inc. (A member of Centaur Plc Group) |
| MortgageStrategy (Online) | Magazines | Centaur Media USA Inc. (A member of Centaur Plc Group) |
| Multinational Monitor | Magazines | Multinational Monitor |
| Nation's Business | Magazines | Chamber of Commerce of the United States |
| Nation's Business (1986-1998) | Magazines | Chamber of Commerce of the United States |
| Nation's Business (pre-1986) | Magazines | Chamber of Commerce of the United States |
| New Jersey Business | Magazines | New Jersey Business & Industry Association |
| Next Step Magazine | Magazines | Next Step |
| Nonprofit World | Magazines | Society for Nonprofit Organizations |
| Northwest Arkansas Business Journal | Magazines | Gray Matters, LLC |
| Oregon Business | Magazines | MEDIAmerica, Inc. |
| PM Network | Magazines | Project Management Institute |
| PPI Detailed Report (Online) | Magazines | Superintendent of Documents, Labor Department |
| Presidents & Prime Ministers | Magazines | EQES, Inc. |
| Regardie's | Magazines | Regardie's |
| Region, The | Magazines | Federal Reserve Bank of Minneapolis |
| Rental | Magazines | AC Business Media |
| Saint Paul Legal Ledger | Newspapers | BridgeTower Media Holding Company |
| Shopper Marketing | Magazines | EnsembleIQ |
| Smart Business Atlanta | Magazines | Smart Business Network |
| Smart Business Broward/Palm Beach | Magazines | Smart Business Network |
| Smart Business Chicago | Magazines | Smart Business Network |
| Smart Business Indianapolis | Magazines | Smart Business Network |
| Smart Business Philadelphia | Magazines | Smart Business Network |
| SMART Manufacturing | Magazines | SME |
| SmartMoney | Magazines | Dow Jones & Company, Inc. Financial Information Services |
| SmartMoney.com | Magazines | Dow Jones & Company, Inc. Financial Information Services |
| Software Executive (Online) | Magazines | Jameson Publishing/Vert Markets |
| St. Charles County Business Record | Newspapers | BridgeTower Media Holding Company |
| St. Louis Commerce Magazine | Magazines | St. Louis Region Commerce and Growth Association |
| Staffing Management | Magazines | Society for Human Resource Management |
| Stanford Social Innovation Review | Magazines | Stanford Social Innovation Review, Stanford University |
| Tax Features | Magazines | Tax Foundation Inc. |
| Transworld Business | Magazines | Source Interlink Companies |
| Vermont Business Magazine | Magazines | Boutin-McQuiston, Inc. |
| Wall Street Journal | Newspapers | Dow Jones & Company Inc |
| Wall Street Journal (Online) | Newspapers | Dow Jones & Company Inc |
| Wall Street Journal Americas, The | Newspapers | Dow Jones & Company Inc |
| Wall Street Journal Asia, The | Newspapers | Dow Jones & Company Inc |
| Wall Street Transcript | Newspapers | Wall Street Transcript |
| Western New York | Magazines | Greater Buffalo Chamber Services Corporation |
| World | Magazines | KPMG Peat Marwick |

Figure A1 shows the frequency of a selective sample of company names in news coverage over time. The peaks in company name mentions trace important business events, suggesting the temporal sensitivity and validity of the business news corpora for our purpose of measuring trends in business association.

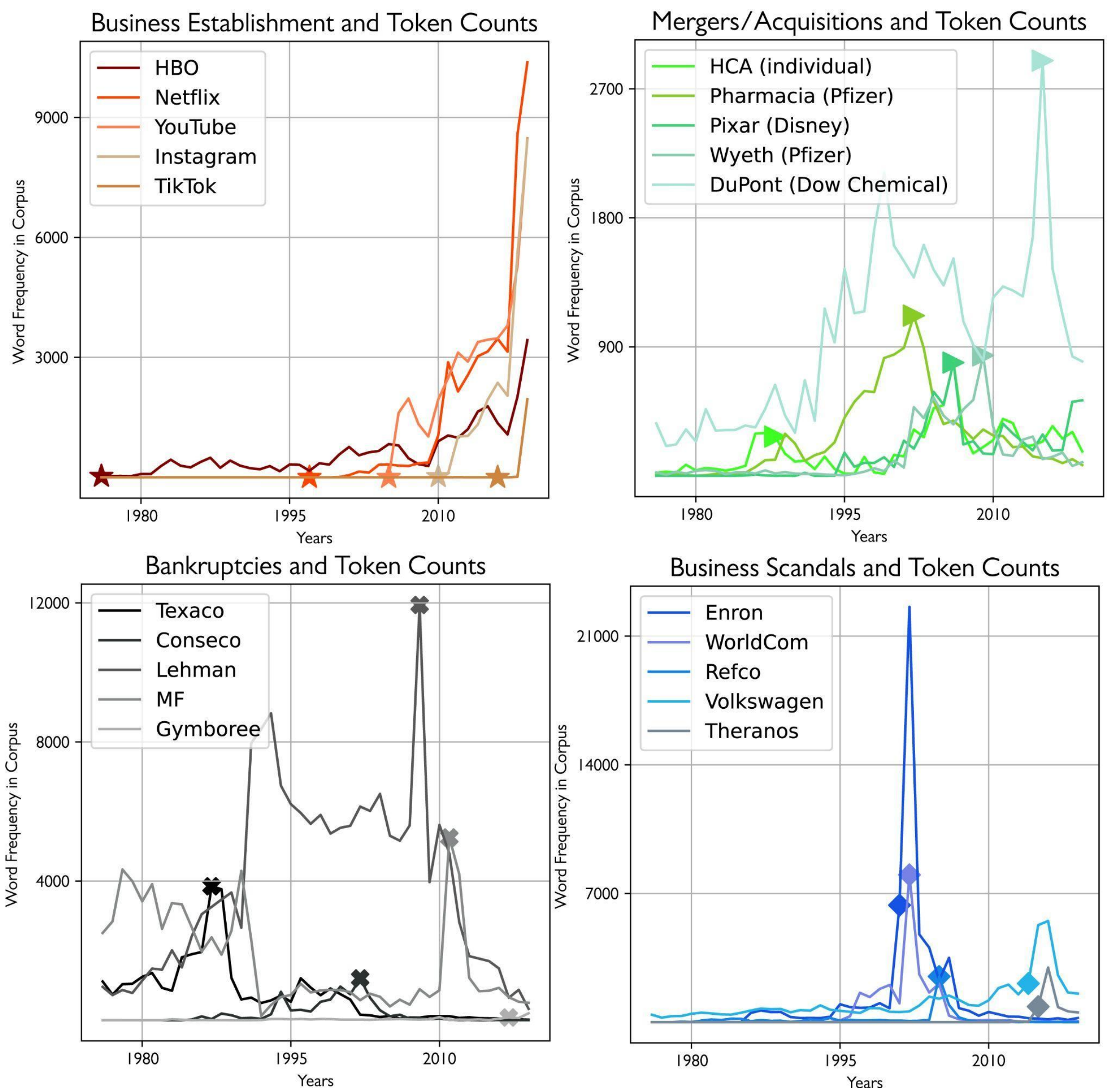

Figure A1. Token Counts of Company Names with Specific Events

## A.2 The Dynamic Word Embedding Space

In this section, we provide additional validation for our dynamic embedding space as a rich trace of historical associations. First we introduced the rationale and computational principles behind dynamic word embedding method; then, we validated this landscape with the performance of the real estate industry in the stock market. Finally, we trace the semantic drift of several important companies, including Apple, Tesla and Theranos; then evaluate all public companies against their stock market valuations.

For word embedding algorithms, the primary embedding output is an *n*-by-*k* matrix, where *n* denotes the number of the words and *k* denotes the word vector dimension.

Distance can be assessed between word vectors, most commonly with cosine distance, to compute semantic relations between their underlying words. A small distance between word vectors reveals close or even identical meanings and associated semantic contexts (e.g., synonyms). By contrast, a large distance in semantic space suggests irrelevance (when the angle between vectors is 90 degrees or more). The precision of geometric distances between word vectors in semantic space also allows analysts to solve analogy problems with simple linear algebra. A well-known example given by (Mikolov et al., 2013a) demonstrates how for most sizable English-language corpora, word vectors $\overrightarrow{king}$ - $\overrightarrow{man}$ + $\overrightarrow{woman}$ from a learned embedding model will be closest to the word vector for $\overrightarrow{queen}$. Precursors to the modern word embedding model relied on matrix factorization approaches, like singular value decomposition (SVD), which constrained the number of words and documents from which distances could be computed. One of the most prominent recent advances, *word2vec* (Mikolov et al., 2013a), heuristically approximates this matrix factorization (Levy and Goldberg, n.d.), but with much greater efficiency. This single advance removes limitations on word and corpus size and allows the analyst to replace documents with small word-windows as relevant semantic contexts, which dramatically increased their accuracy.[9]

Approaches to embedding alignment and temporal comparison have become more sophisticated in recent years, giving birth to the family of algorithms commonly known as "dynamic word embedding" method (Hamilton et al., 2016; Liu et al., 2020; Yao et al., 2018; Zhang et al., 2016). Outputs from these algorithms are time-stamped word vectors that contain the semantic information for that specific period and are comparable across history. Dynamic word embedding algorithms tend to follow one of two strategies. The first approach typically aligns time-slice embeddings by constructing a linear transformation between them, assuming that most words do not change in meaning across most time periods. This has been achieved through solving a $d$-dimensional least-squares problem to align each word and its nearest neighbor words across time-adjacent embeddings (Kulkarni et al., 2015), as through the Procrustes procedure (Hamilton et al., 2016). Some analysts have further stipulated that the alignment be calculated for "anchor words" of comparable high frequency in all time periods (Zhang et al., 2016). The challenge with these algorithms is that they all require overlapping words across embeddings, and drop new vocabulary introduced into the cultural system over time. For technology and business, where disruptive innovation often occurs through the emergence of new components, this can lead to inaccurate innovation estimates as it systematically ignores all elements not existing in the first time period. In this work, we adopt the joint training paradigm in order to naturally account for changes in the composition of words used over historical time (Yao et al., 2018).

[9] Several other approaches have built on this basic architecture (Joulin et al., 2016; Pennington et al., 2014). More recent embeddings built from transformers allow us to calculate different word vectors for each context in which the word appears (Devlin et al., 2018). These models, however, have from hundreds of millions to trillions of parameters and require massive web corpora in order to pre-train them. Prior research has demonstrated that non-contextual word embeddings are more accurate than tuned contextual ones when the analyzed corpus is sufficiently different from the pre-trained one ([self-citation placeholder]), and so we used dynamic non-contextual models in this analysis.

The output from word embedding space forms a robust foundation for calculating distance and relationship between social entities. As the dimensionality of a hypersphere grows, its volume shrinks relative to its surface area as more of that volume resides near the surface. In our 50 dimensional word embedding, all normalized word vectors lie on the surface of a 50 dimensional hypersphere (Levy et al., 2015) such that orthogonality or unrelatedness is indicated by a cosine distance of one, equivalent to an angle of 90°. When the angle between words shrinks from 90°, it suggests a meaningful relationship between them.[10]

How does our embedding model perform? We validate this landscape with the performance of the real estate industry in the stock market. We first build a dimension of profit or loss, based on the method from precedent (Kozlowski et al., 2019). The positive words include “gain”, “win”, “profit”, “bull”, “optimistic”, “worthy” and “profitable”, while the negative words include “lose”, “loss”, “default”, “bear”, “pessimistic”, “worthless” and “unprofitable”. We project the vectors for words, “estate”, “homeowner”, “house”, “real estate” and “mortgage”, onto this dimension. The profiting score from this vector projection algorithm shows the market emotion to the real estate industry, based on news coverage from major business publications. Results show a high correlation between this projection series and the NAHB housing market index, a ground truth index based on surveys collected by the national association of home builders. The two series are cointegrated (with t=-4.0025 and p=0.007 in Engle-Granger two-step cointegration test). For 2008, we can see a collapse of the real estate industry, which leads to a systematic crisis in the financial market (and collapse of the S&P 500 index; see Figure A2.)

[10] We know this from the Gaussian Annulus theorem, that two random points from a d-dimensional Gaussian with unit variance in each direction are approximately orthogonal (Blum, Hopcroft, and Kannan 2016). Semantic relatedness can be inferred from angles less than 90°, but not those greater because the model is trained on empirically proximate words relative to random, unrelated words, not antonyms with a “negative” semantic meaning.

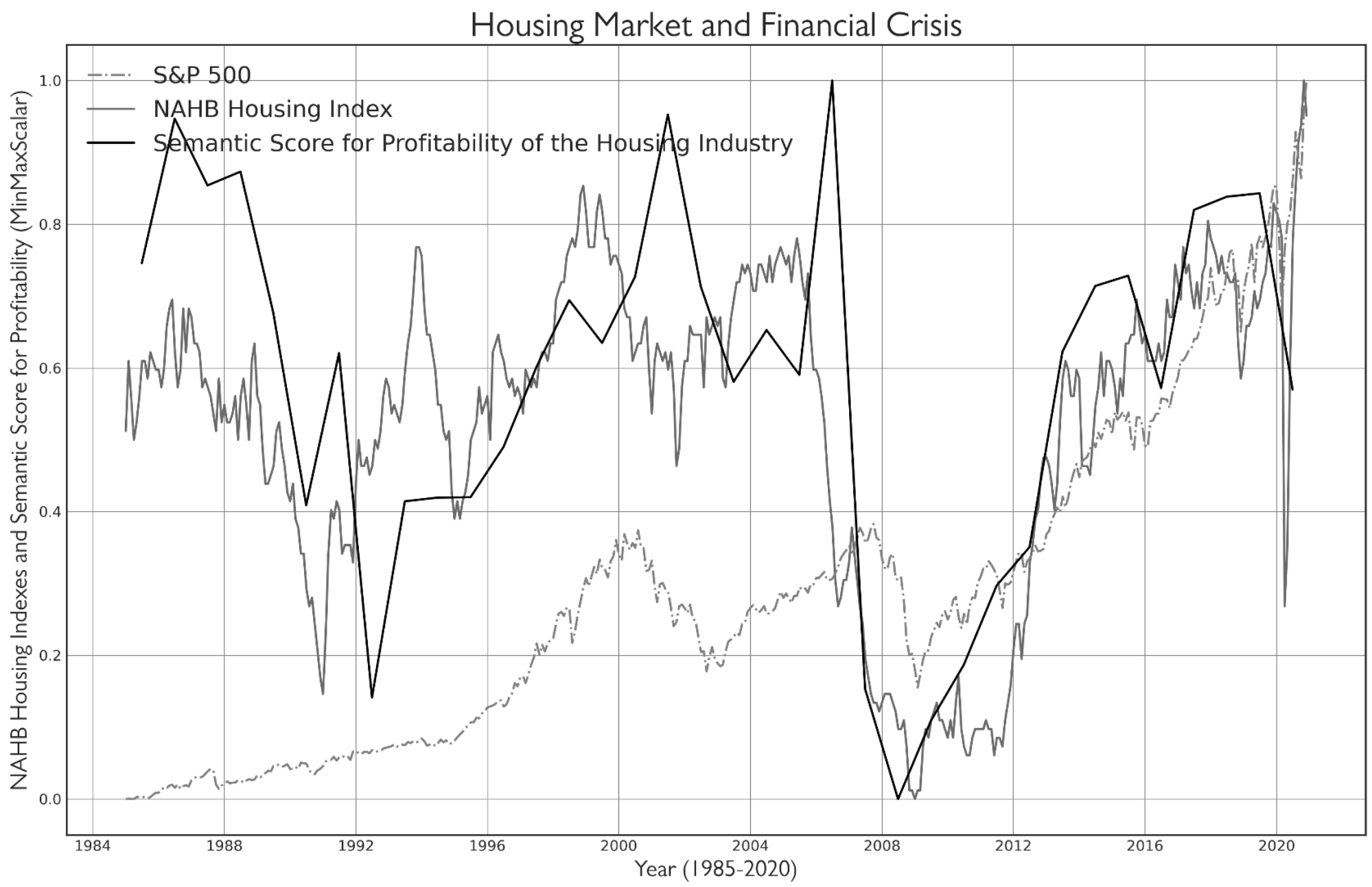

Figure A2. Validation of Dynamic Word Embedding with Stock Market data.

Finally, we trace the semantic drift of several important companies. Figure A3 tracks drifting semantic pathways for several company names across time. Locations of company names in semantic space are determined by their publicly discussed strategies, events, and other subjective impressions. We adopt the same method as Figure 4, described in the main text. Results are largely intuitive. Between 1976 and 2020, Apple transformed from fruit to a prestigious IT company. Tesla was initially characterized as a 19th-century engineer but emerged as an electric car producer. Theranos began with a positive public image, but became linked with other stigmatized companies like Enron and with legislative and judicial actions like “scrutinize” as it became embroiled in scandal.

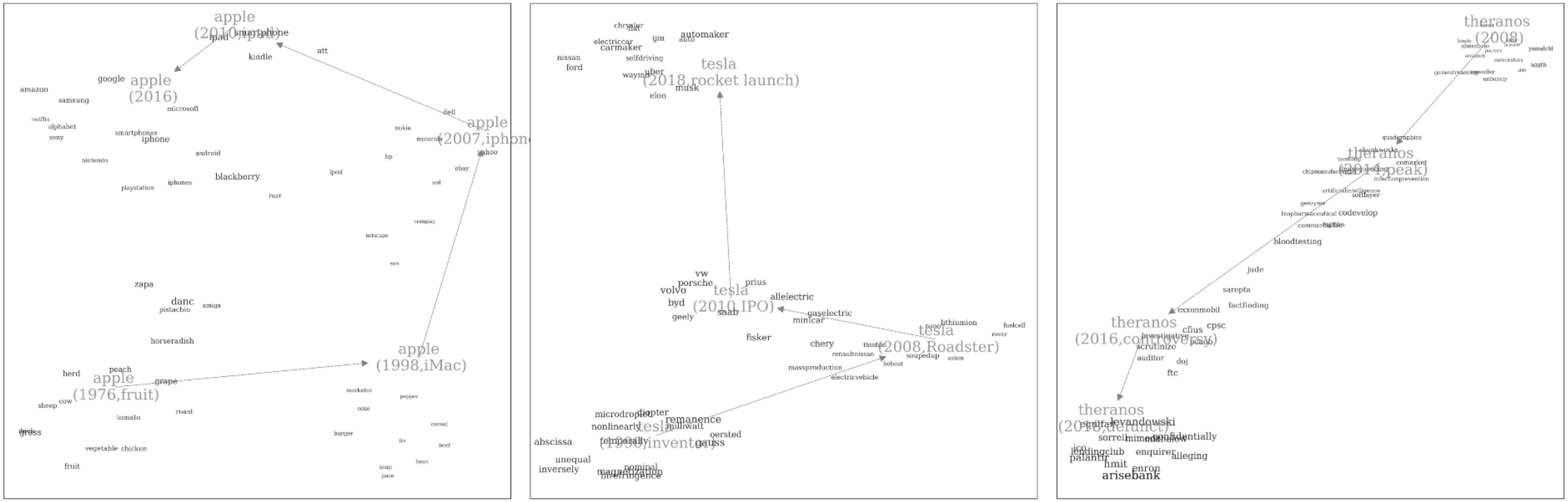

Figure A3. U-MAP Visualization of the Business Dynamic Traces. Left to Right: Apple; Tesla; Theranos.

In a second validation, we focus on public companies present in the CrunchBase dataset. For each pair of public companies, we measure (1) their closeness in semantic space by calculating cosine similarity between their description projections, (2) the correlation between their stock prices for overlapping periods in which they were publicly listed. In total, we have 3,346 valid companies and 4,972,430 distinct pairs.

We cut semantic proximity and stock price correlation coefficient into 20 identical percentile groups, and visualize the 2020 heatmap in Figure A4. We present the co-variation of stock prices across industries in Figure A5, where node size is proportional to intra-industry average correlation, and width/color of the line represents the inter-industry correlation.

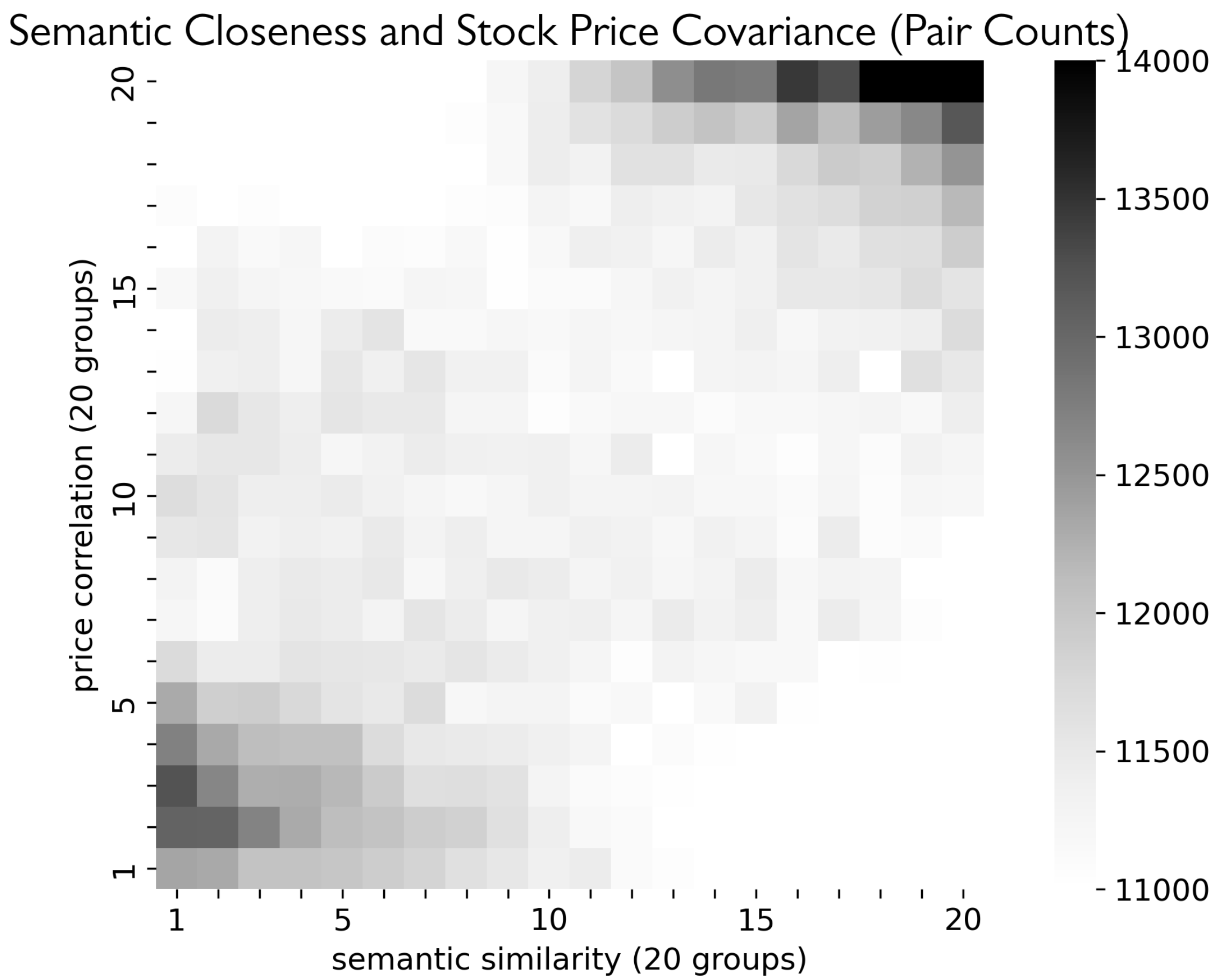

Figure A4. Semantic Closeness and Stock Price Co-Variation

Figure A4 shows a positive relationship between semantic closeness and financial closeness. When two companies are close in semantic space, and thus similar in described business model, market niche, and strategy, their stock prices tend to move in a similar direction. This might be accounted for as they become influenced by similar macro- and microeconomic factors.

In Figure A5, industries including financial services, tourism, real estate, events, and entertainment, etc. are in the center of the stock co-movement network. These are the industries influenced disproportionately by common macro-economic factors.

We also test the relationship between semantic closeness and financial co-variation at the industrial level. The correlation coefficient is approximately -0.33, which reveals that a more semantically homogeneous industry tends to manifest fiercer competition, yielding a higher divergence between stock prices.

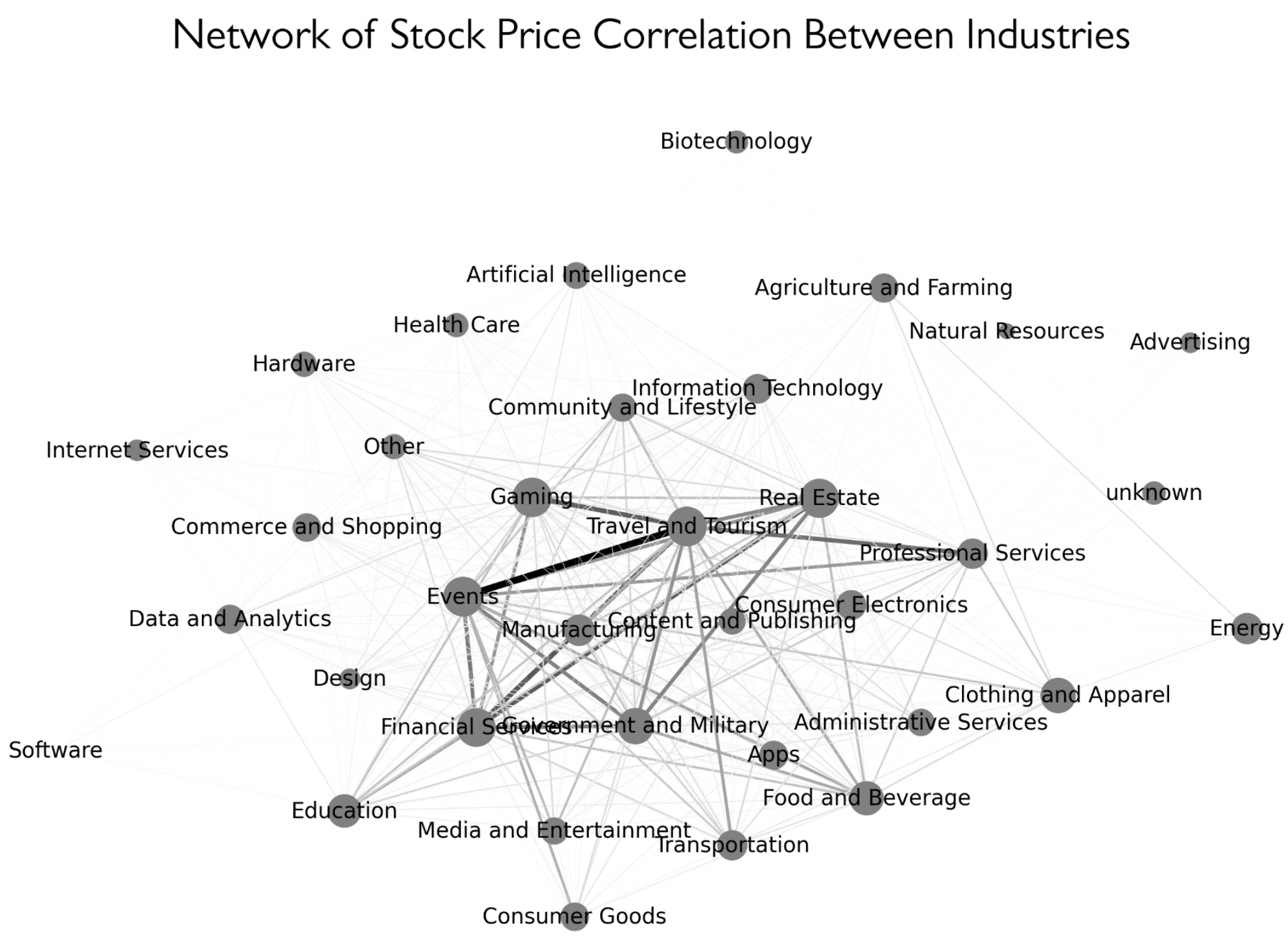

Figure A5. Network of Stock Price Correlation Between Industries

# Appendix B. Examples of Business Descriptions and the Modular Structure of Innovation

To provide a clear display of modular structure of recombination, consider the firms PayBito, a trading platform for cryptocurrencies, and ProNeurotech, a developer of neuroprotective drugs. The descriptions of PayBito and ProNeurotech are listed as below:

***PayBito:*** *PayBito is a bitcoin and cryptocurrency exchange trading platform for major cryptocurrencies including Bitcoin, Ether, Bitcoin cash, Litecoin, HCX, Ripple, ECR20 and many more. The platform allows purchases in INR, USD and supports cryptocurrency trading pairs. Available in Web version, Android & iOS App.*

***ProNeurotech:*** *ProNeurotech is a developer of neuroprotective drugs that prevent axon loss after acute injury or chronic degenerative disease. The company's business includes research in biochemistry, genetics, neurology and synthetic chemistry that drive the synthesis of the cardinal metabolite, nicotinamide adenine dinucleotide, enabling patients with neurodegenerative diseases to have access to treatment of neurologic and ocular diseases.*

Both firms are innovative in the sense that they contain some technical components and draw together previously remote domains. We visualize their semantic elements, in the year of their establishment, projected to 2 dimensions as the empirical landscape with U-Map visualization as in Figure B1.

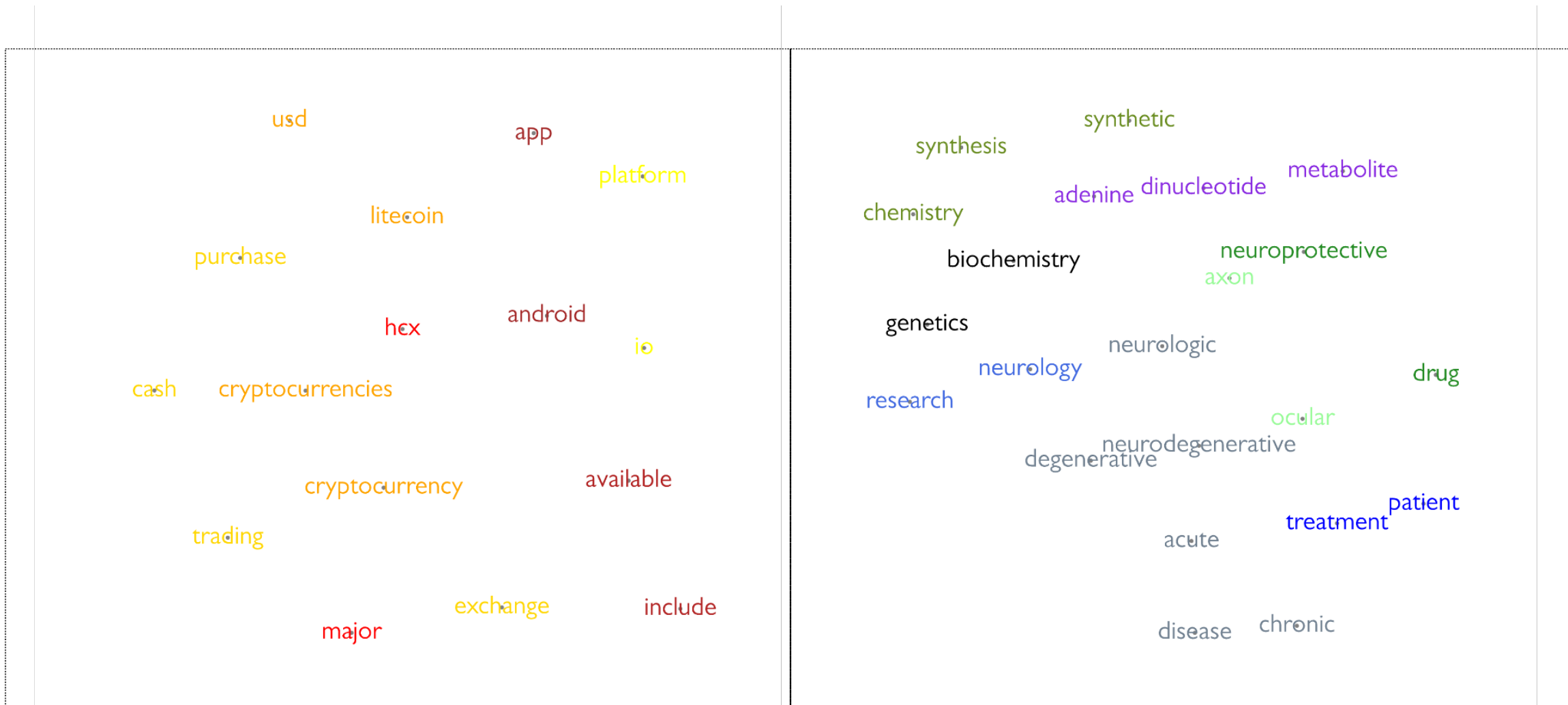

Figure B1. PayBito and ProNeurotech in the Year of Establishment.*Left: PayBito; Right: ProNeurotech.

Each of the colors shows that the elements belong to a certain topic, based on discourse atom modeling (DAT). The figures provide an intuitive way to understand the combinatorial nature of new venture strategies: PayBito aligns internet platforms with cryptocurrency trading, while ProNeurotech combines neurology with synthetic chemistry to solve challenges associated with axon loss.

# Appendix C. Robustness Tests with Alternative Methods for Variable Construction

Both our computational modeling and conclusions depend on the accurate identification of the modular structure of the economic system. In our original analysis, we use discourse atom modeling to divide the space into 200 sections as the modular structure to support innovative strategies. Here, we (1) change the total number of discourse atoms between 200 and 100; (2) change the space division technique from discourse atom modeling (DA, 200 atoms) to k-means clustering (KM, 200 clusters); and and (3) either exclude isolated modules with only a few tokens or retain them in the calculation (FULL). After identifying the modular structure using these alternatives, we recalculate the variables based on the new module definitions, compare the measures (Figure C1), and replicate the regression analysis (Table C2). The results remain largely consistent, confirming the robustness of our conclusions.

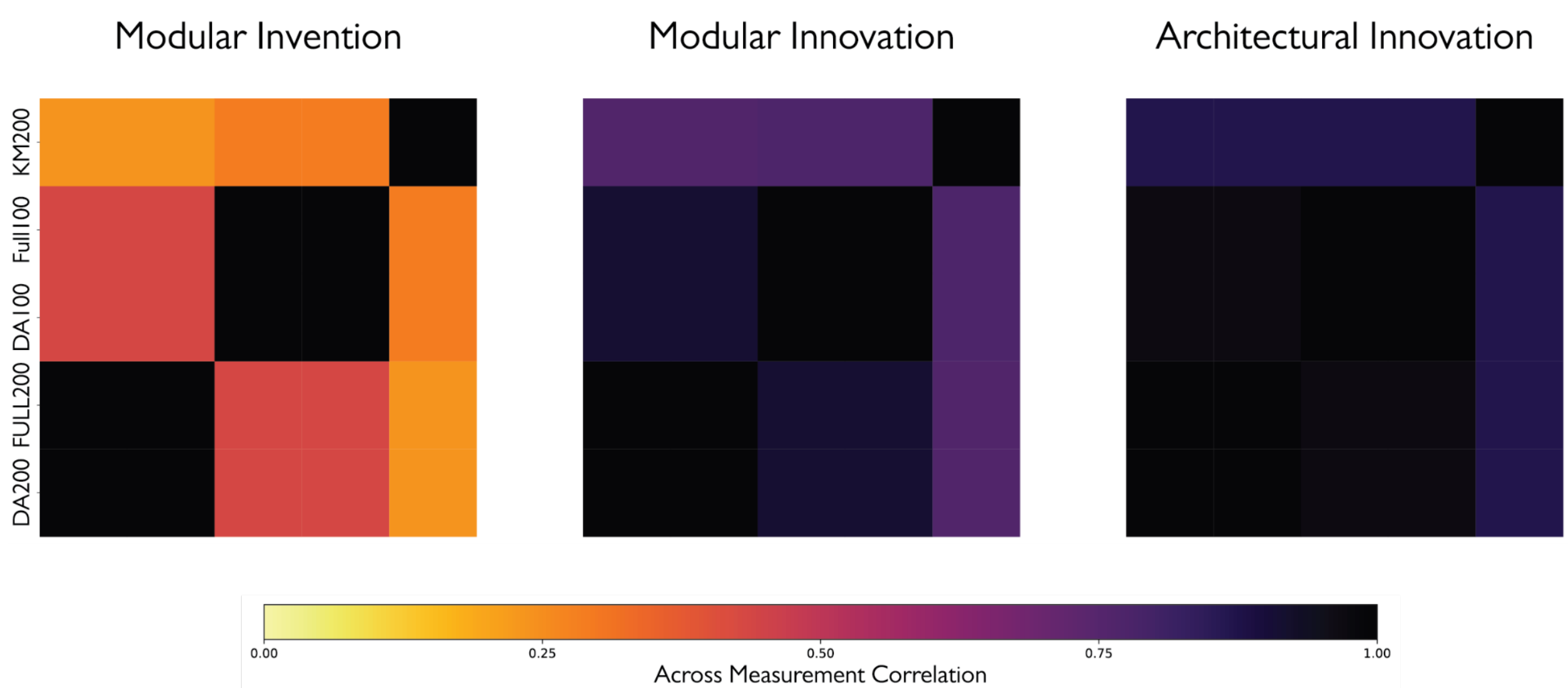

Figure C1. Correlation between Major Variables based on Alternative Methods.

Figure C1 presents the correlations among five sets of measures generated using alternative methods. Both architectural and modular innovation measures show strong correlations (coefficients well above 0.75). For modular invention, KMeans clustering yields slightly different values, but overall the variables are only minimally affected by alternative module identification and remain highly correlated across methods. The results in Table C2 are broadly consistent with those in Table 4 and Figure 6, reinforcing the conclusion that new ventures engaging in architectural innovation are more likely to succeed in competition.

Table C2. FPM Survival Models with Alternative Module Identification

| | **Alternative: Discourse Atom (100)** | | | | **Alternative: KMEANS Clustering (200)** | | | |
|---|---|---|---|---|---|---|---|---|
| | IPO/High Price Acquisition | New Funding | Other Acquisition | Close | IPO/High Price Acquisition | New Funding | Other Acquisition | Close |
| Local Distance (Modular Invention) | -0.855*** | -0.444*** | 0.492*** | 0.353** | -0.912*** | -0.104** | 0.220*** | 0.087 |
| | (0.164) | (0.033) | (0.045) | (0.135) | (0.178) | (0.039) | (0.057) | (0.165) |
| Global STD (Modular Innovation) | -1.117** | -0.879*** | -0.501*** | 1.787*** | -0.155 | -0.605*** | -0.211 | 1.260*** |
| | (0.384) | (0.082) | (0.147) | (0.381) | (0.301) | (0.066) | (0.124) | (0.329) |
| Global Distance (Architectural Innovation) | 2.363*** | 1.014*** | 0.026 | -2.626*** | 1.443*** | 0.600*** | 0.009 | -2.180*** |
| | (0.292) | (0.061) | (0.100) | (0.261) | (0.235) | (0.052) | (0.090) | (0.246) |
| Elements Familiarity | -0.156*** | 0.025*** | 0.060*** | -0.063*** | -0.158*** | 0.022*** | 0.058*** | -0.062*** |
| | (0.016) | (0.002) | (0.004) | (0.010) | (0.016) | (0.002) | (0.004) | (0.010) |
| Dummy: Usage of social media | -0.062* | 0.173*** | -0.403*** | -0.397*** | -0.071** | 0.176*** | -0.406*** | -0.408*** |
| | (0.027) | (0.006) | (0.012) | (0.030) | (0.027) | (0.006) | (0.012) | (0.030) |
| Dummy: founded by women or minorities | -0.546*** | 0.292*** | -0.441*** | -0.268*** | -0.515*** | 0.297*** | -0.433*** | -0.275*** |
| | (0.055) | (0.007) | (0.026) | (0.044) | (0.055) | (0.007) | (0.026) | (0.044) |
| Market growth rate in the year of establishment | -0.050 | 0.006 | 0.030 | -0.128 | -0.045 | 0.007 | 0.037 | -0.132 |
| | (0.067) | (0.015) | (0.030) | (0.088) | (0.066) | (0.015) | (0.030) | (0.088) |
| Text Features (length, missing, unusual words) | YES | YES | YES | YES | YES | YES | YES | YES |
| Year, Location, Industry and Funding History | YES | YES | YES | YES | YES | YES | YES | YES |
| Constant | -5.024*** | -4.280*** | -4.705*** | -6.464*** | -4.676*** | -4.177*** | -4.689*** | -6.555*** |
| | (0.145) | (0.062) | (0.059) | (0.355) | (0.124) | (0.060) | (0.057) | (0.353) |

*Note: ***p < .001, **p < .01, *p < .05, two-tailed test.*

# Appendix D. Robustness Tests with Alternative Models and Datasets

## D.1. Validation with Cause-Specific Cox Model

The cause-specific cox model is an alternative, commonly-used strategy to estimate the hazard rate for competing events. Given its assumption that competing events are independent, it remains widely used and has been accepted as the default model within the social science community. We re-run models in Table 4 with the cause-specific Cox model, with results reported below in Table D1.

Table D1. Replication of Event History Analysis with Cause Specific Cox Model

| | IPO/high price acquisition | Getting New funding | Other Acquisition | Close |
|---|---|---|---|---|
| Local Distance (Modular Invention) | -1.011*** | -0.409*** | 0.411*** | 0.064 |
| | (0.149) | (0.030) | (0.050) | (0.125) |
| Global STD (Modular Innovation) | -1.424*** | -1.285*** | -1.774*** | 1.235** |
| | (0.411) | (0.091) | (0.167) | (0.402) |
| Global Distance (Architectural Innovation) | 3.026*** | 1.227*** | 0.768*** | -2.435*** |
| | (0.319) | (0.069) | (0.120) | (0.267) |
| Technology Elements Familiarity | -0.194*** | 0.029*** | 0.101*** | -0.034** |
| | (0.019) | (0.002) | (0.006) | (0.011) |
| Dummy: Usage of social media | -0.159*** | 0.040*** | -0.516*** | -0.563*** |
| | (0.027) | (0.007) | (0.012) | (0.030) |
| Dummy: founded by women or minorities | -0.383*** | 0.286*** | -0.296*** | -0.085 |
| | (0.062) | (0.007) | (0.028) | (0.045) |
| Market growth rate in the year of establishment | 0.086 | 0.077*** | 0.070* | 0.050 |
| | (0.065) | (0.016) | (0.032) | (0.089) |
| Text Features (length, missing, unusual words) | YES | YES | YES | YES |
| Year, Location, Industry and Funding History | YES | YES | YES | YES |
| Goodness of Fit ($\rho_k^2$) | 0.501 | 0.405 | 0.468 | 0.763 |
| Goodness of Fit (R2) | 0.379 | 0.292 | 0.348 | 0.662 |

*Note: ***p < .001, **p < .01, *p < .05, two-tailed test.*

Results of the cause-specific Cox model are close to the results of FPM. Nearly all of the effects from our invention measures on four types of outcome events are replicated, except that higher-order invention becomes positively related to other acquisition events when competing risks are unconsidered. The model shows the consistency of our conclusion across different empirical methods.

## D.2. VentureXpert Validation

We use another independent dataset, VentureXpert, to repeat our model in Table D2. VentureXpert was developed by Thomson One, which provides comprehensive data about market quotes, earnings estimates, financial fundamentals, transaction data, corporate filings, ownership profiles, etc., for major business organizations in the U.S. We collected the VentureXpert dataset in 2018, and tried to include all start-ups established in the U.S. after 1980. We obtained a sample of 63,492.

In the sample of company overlaps between the Crunchbase and VentureXpert databases, descriptions are completely different for the same company—they are not copied by analysts from Crunchbase descriptions. Nevertheless, the cosine distances within Crunchbase and VentureXpert descriptions for the same companies correlate at 0.2, $p<.001$. When we calculate the distance between each pair of matched ventures in the space, and compare them with a random baseline (distance between random pair of ventures), a $t$-test shows that the matched pairs have significantly lower distances between them ($t$ = -124 with $p<0.001$). This suggests that even though the descriptions are independently constructed, they reference similar underlying concepts. Because of their independence in source of description, but also massive differences in samples, we analyze the two datasets separately as semi-independent empirical validations of all our hypotheses above.

For comparison between Crunchbase and VentureXpert datasets, we visualize the location of each industry within the two industry classification systems, calculated as the average location of all ventures within the industry. While VentureXpert classifies the new ventures with standard SIC codes, Crunchbase classifies them with their own industry categories (for details, turn to the Crunchbase support page at [https://support.crunchbase.com/hc/en-us/articles/360043146954-What-Industries-are-included-in-Crunchbase-](https://support.crunchbase.com/hc/en-us/articles/360043146954-What-Industries-are-included-in-Crunchbase-)). To rule out year heterogeneity, we calculate the 2020 location for all ventures. Results from this industry comparison are displayed in Figure A8. The center of bubbles shows the position of each industry, while the size of bubbles shows the number of ventures.

Figure D1 shows high consistency between Crunchbase and VentureXpert industry locations. The "Manufacturing" category is nearly overlapping. VentureXpert "Mining" is extremely close to Crunchbase "Natural Resources", "Energy" and "Sustainability"; VentureXpert "Agriculture,

Forestry and Fishing" is very close to Crunchbase "Agriculture and Farming", while VentureXpert "Finance, Insurance and Real Estate" is co-located with Crunchbase's "Lending and Investment", "Financial Services" and "Payments". Other categories, although not mentioned here, also correspond intuitively.

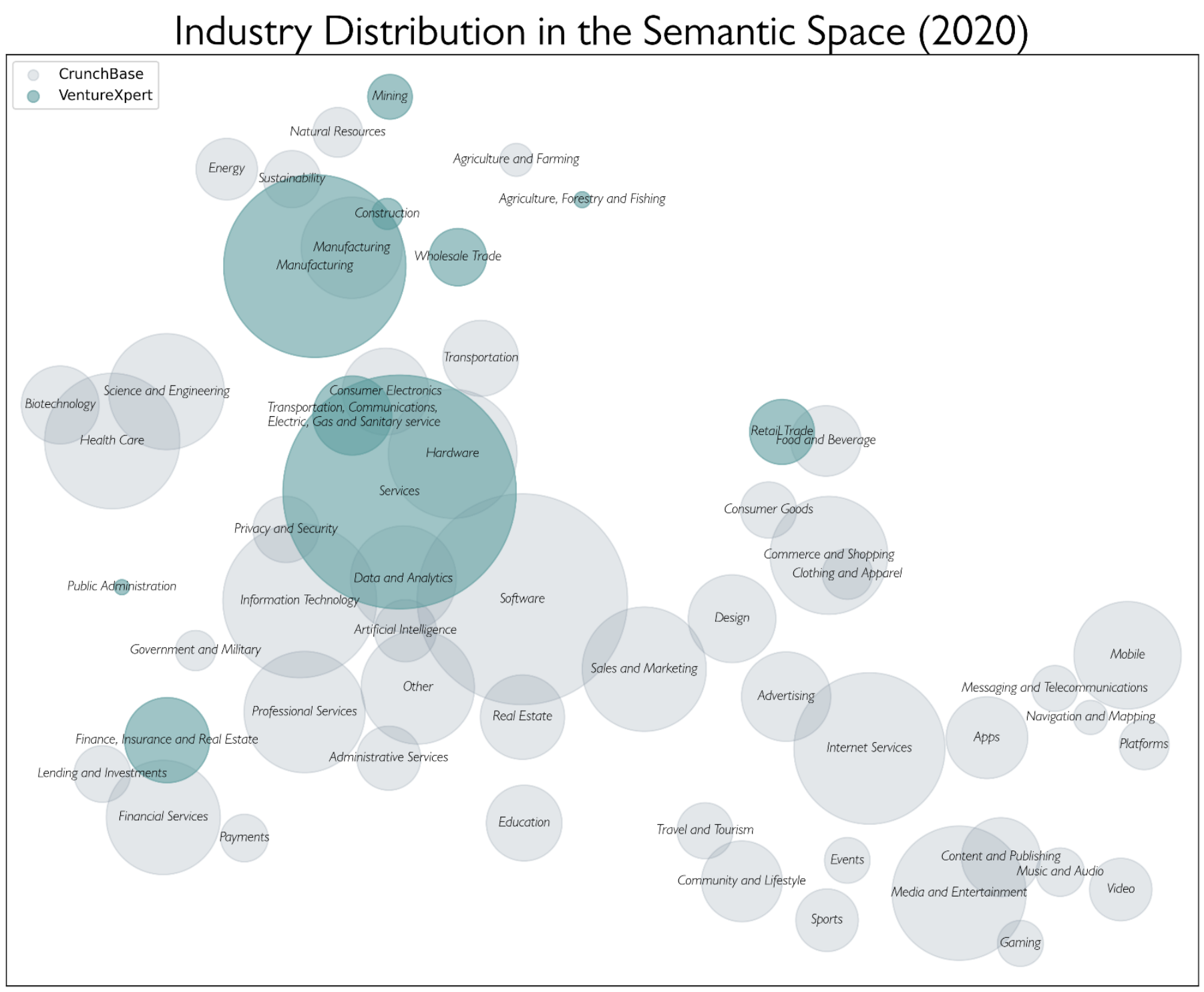

**Figure D1. Comparison between Crunchbase and VentureXpert Industry Locations**

Next, we replicate our event history model on VentureXpert. As the data collection process and data structure of VentureXpert is different from CrunchBase, we carried out some preprocessing procedures to make the distribution of the two datasets comparable. First, as we do not have the exact time for bankruptcies in VentureXpert, we delete all samples that have any bankruptcy event in both datasets simultaneously in the following comparison. Second, as VentureXpert has a much higher proportion of funding events compared with CrunchBase (see table 2), we randomly recode 50% of reported funding events in VentureXpert as censored, to keep proportions close. Finally, the two datasets have different outcome distributions in different year groups, which implies potential bias in data collection. For each dataset, we generate a weight equal to the whole sample outcome distribution, respectively, and use this weight in model estimation.

The comparison models for the two datasets are reported in Table D2, and the hazard rate prediction figure is visualized in Figure D2. The outcome shows that the same pattern appears in the two datasets, and our results are robust across contexts.

Table D2. Comparative FPM for CrunchBase and VentureXpert

| | IPO/RTO | Getting new funding | Acquisitions (All) |
|---|---|---|---|
| **CrunchBase** | | | |
| Local Distance (Modular Invention) | -0.994*** | -0.451*** | 0.519*** |
| | (0.168) | (0.029) | (0.042) |
| Global STD (Modular Innovation) | -0.358 | -0.970*** | -0.397** |
| | (0.435) | (0.084) | (0.148) |
| Global Distance (Architectural Innovation) | 2.431*** | 1.130*** | -0.103 |
| | (0.336) | (0.063) | (0.101) |
| Other Variables | | Controlled | |
| Goodness of Fit | | $\rho_k^2$=0.403, R2=0.291 | |
| **VentureXpert** | | | |
| Local Distance (Modular Invention) | -0.641* | -0.447*** | -0.521*** |
| | (0.324) | (0.051) | (0.112) |
| Global STD (Modular Innovation) | -0.431 | -0.076 | -0.362 |
| | (0.714) | (0.132) | (0.300) |
| Global Distance (Architectural Innovation) | 2.143*** | 0.327*** | 0.205 |
| | (0.481) | (0.091) | (0.210) |
| Other Variables | | Controlled | |
| Goodness of Fit | | $\rho_k^2$=0.181, R2=0.119 | |

*Note: ***p < .001, **p < .01, *p < .05, two-tailed test.*

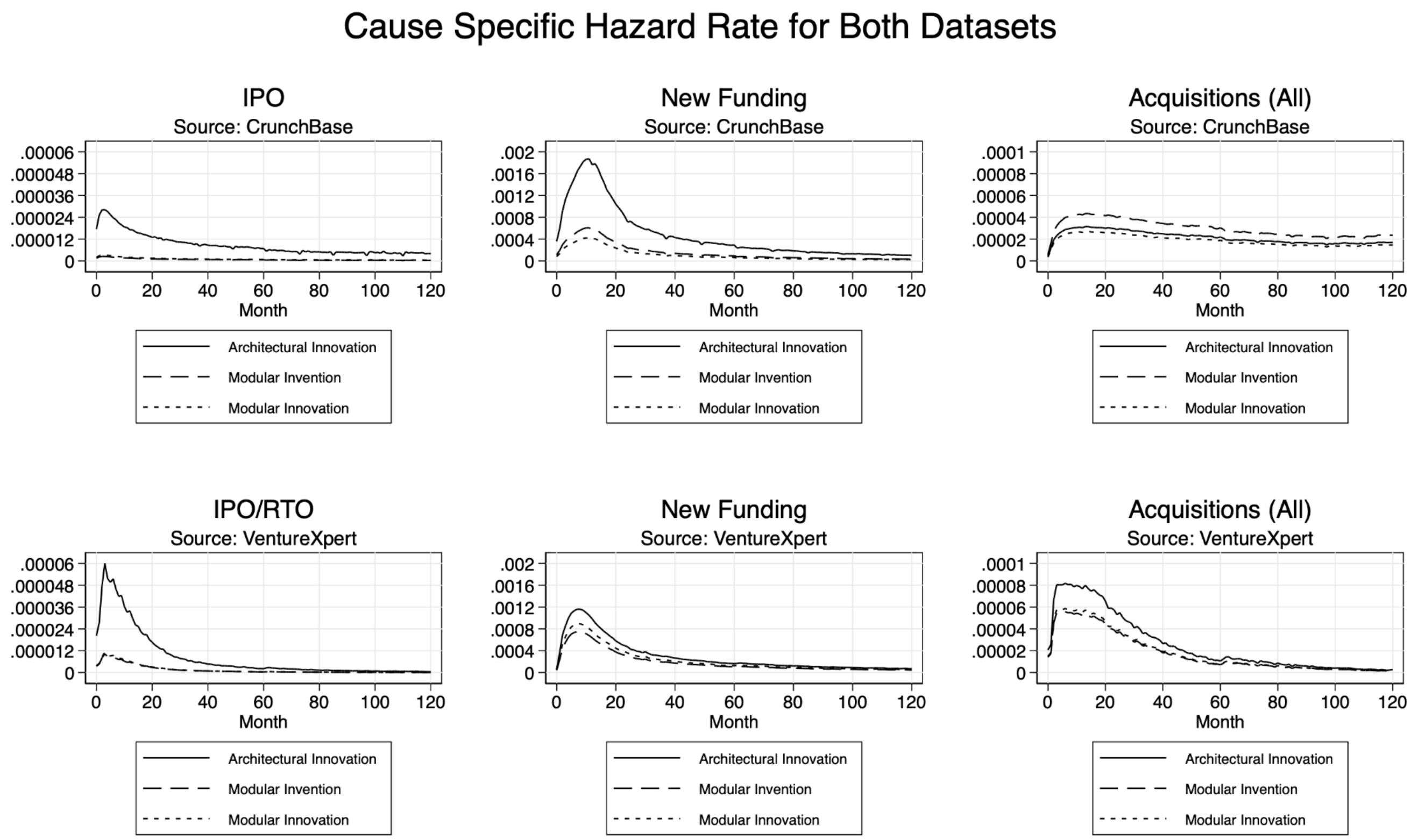

**Figure D2. Hazard Rate Prediction Comparison for Both Models**

## D.3. Wayback Machine Validation

To account for time-varying changes in strategy for start-ups, we turned to the Wayback Machine and scraped historical records from CrunchBase. The Wayback Machine is a non-profit digital archive of the internet, which scrapes the whole net frequently to archive a snapshot that allows users to “go back in time” and see how the website looks in the past. Here, for companies with several records in one year, we only keep one record each year to reduce redundancy. In total, we have approximately 150,000 companies that have at least two historical records between 2013 and 2021 in our dataset, which makes over 50% of the whole sample. Among them, 58,437 samples provide records before 2018. In this validation, we calculate measures of recombination based on the following contingency: if the time point of interest is out of the recorded period, we use the closest description record; if the time point of interest falls between two valid records, we calculate the value based on an assumption of even transformation speed and smoothly interpolate between the two measures. A visualized explanation of this calculation is shown in Figure D3.

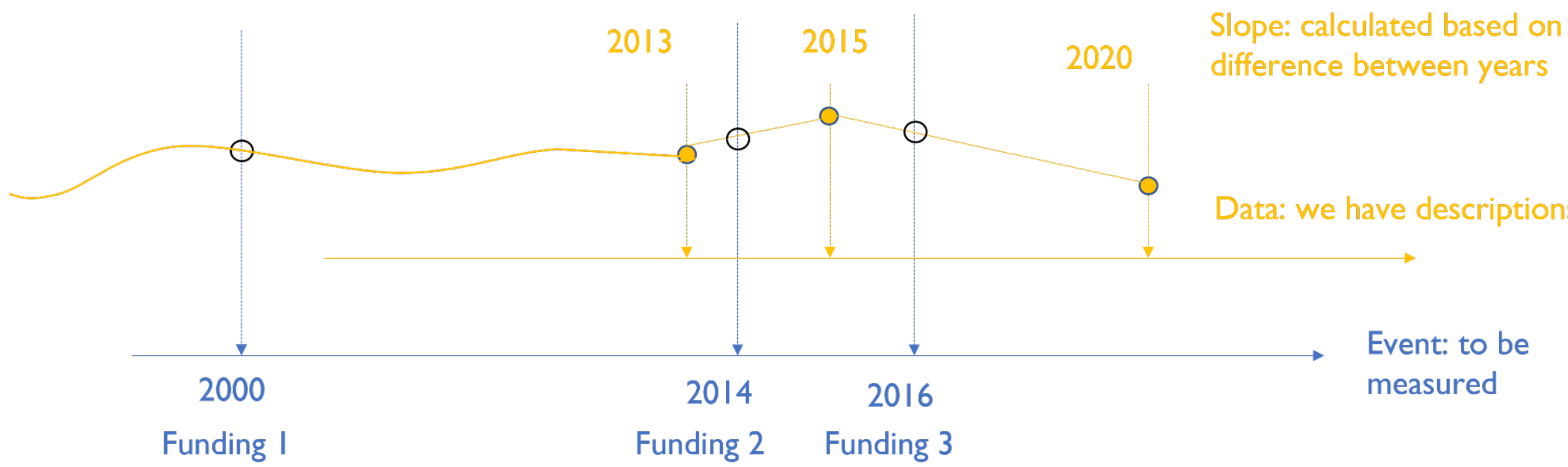

**Figure D3. Conceptual Figure for Wayback Machine Measurement Construction**

After we obtain these measures, we repeat the model in Table D3. Results in Table D3 show that our basic conclusions remain robust, although some significant changes occur due to the variance in sample size.

Table D3. The Wayback Machine Based Validation of FPM

| | IPO/high price acquisition | Getting New funding | Other Acquisition | Close |
|---|---|---|---|---|
| Local Distance (Modular Invention) | 1.253* | -0.861*** | 0.820*** | -0.482 |
| | (0.509) | (0.119) | (0.226) | (0.617) |
| Global STD (Modular Innovation) | 0.706 | -1.065*** | -0.688** | 0.877 |
| | (0.512) | (0.113) | (0.217) | (0.565) |
| Global Distance (Architectural Innovation) | 2.318*** | 0.984*** | -0.025 | -3.594*** |
| | (0.378) | (0.078) | (0.159) | (0.364) |
| Technological Element Familiarity | -0.230*** | 0.018*** | 0.096*** | -0.057*** |
| | (0.025) | (0.003) | (0.007) | (0.016) |
| Dummy: Usage of social media | -0.131*** | 0.170*** | -0.404*** | -0.383*** |
| | (0.032) | (0.008) | (0.015) | (0.038) |
| Dummy: founded by women or minorities | -0.795*** | 0.323*** | -0.597*** | -0.243*** |
| | (0.083) | (0.009) | (0.036) | (0.059) |
| Market growth rate in the year of establishment | 0.021 | 0.009 | 0.079* | -0.137 |
| | (0.081) | (0.020) | (0.039) | (0.115) |
| Constant | -5.274*** | -4.119*** | -4.504*** | -5.625*** |
| | (0.190) | (0.074) | (0.087) | (0.446) |
| Text Features (length, missing, unusual words) | YES | YES | YES | YES |
| Year, Location, Industry and Funding History | YES | YES | YES | YES |
| Goodness of Fit | $\rho_k^2$ =0.140, R2=0.090 | | | |

*Note: ***p < .001, **p < .01, *p < .05, two-tailed test. N* of organizations=58437. *N* of event records=76810.

## D.4. Validation with Alternative Coding of Dependent Variables

In our main text, we consider both IPOs and high-priced acquisitions as milestones of new venture success, as suggested by prior research. However, due to significant missing data in Crunchbase regarding acquisition prices, here the identification of high price acquisition may suffer from low recall. To address this concern, we explore two alternative approaches for coding dependent variables: either (1) treating high-priced acquisitions as a single category or (2) combining high-priced acquisitions with other types of acquisitions, without emphasizing their sale prices. We then re-run our primary model using these two alternative coding methods. The results for this robustness check are reported in Table D4.

From table D4, we can see that changing the coding schema does not change our main results, and our conclusions remain robust.

Table D4. Effects for Main Independent Variables by Alternative Coding

| | Alternative Coding 1: Using High Price Acquisition as Single Category | | | | |
|---|---|---|---|---|---|
| | *IPO* | *High Price Acquisition* | *New Funding* | *Other Acquisitions* | *Close* |
| Local Distance | -1.058*** | -0.634* | -0.424*** | 0.520*** | 0.156 |
| | (0.168) | (0.301) | (0.029) | (0.043) | (0.123) |
| Global STD | -0.687 | -0.066 | -0.947*** | -0.469** | 2.178*** |
| | (0.438) | (0.873) | (0.084) | (0.152) | (0.387) |
| Global Distance | 2.746*** | 0.490 | 1.123*** | 0.009 | -2.837*** |
| | (0.342) | (0.660) | (0.063) | (0.103) | (0.261) |
| Other Variables | YES | YES | YES | YES | YES |
| Goodness of Fit | | | $\rho_k^2$=0.400, R2=0.289 | | |
| | **Alternative Coding 2: Combining All Acquisitions as Single Category** | | | | |
| | *IPO* | | *New Funding* | *All Acquisitions* | *Close* |
| Local Distance | -1.065*** | | -0.425*** | 0.484*** | 0.157 |
| | (0.168) | | (0.028) | (0.042) | (0.123) |
| | -0.681 | | -0.944*** | -0.457** | 2.175*** |
| | (0.439) | | (0.084) | (0.148) | (0.387) |
| Global Distance | 2.752*** | | 1.119*** | 0.003 | -2.836*** |
| | (0.342) | | (0.063) | (0.100) | (0.261) |
| Other Variables | YES | YES | YES | YES | YES |
| Goodness of Fit | | | $\rho_k^2$=0.399, R2=0.288 | | |

*Note: ***p < .001, **p < .01, *p < .05, two-tailed test.*